\documentclass[superscriptaddress,twocolumn,aps,pra,preprintnumbers,notitlepage,showpacs,nofootinbib]{revtex4-1}
\usepackage{amsmath,amssymb}
\usepackage{lipsum} 
\usepackage[latin9]{inputenc}
\usepackage{graphicx}
\usepackage{dcolumn}
\usepackage{bbold}
\usepackage{bm}
\usepackage[usenames]{xcolor}
\usepackage[colorlinks,bookmarks=false,citecolor=blue,linkcolor=blue,urlcolor=blue,hyperfootnotes=true]{hyperref}
\usepackage[caption=false,labelformat=simple]{subfig}
\usepackage{accents}

\begin{document}

\title{Edge magnetization and spin transport in an SU(2)-symmetric Kitaev spin liquid}
\author{V. S. de Carvalho}
\email{vsilva@ifi.unicamp.br}
\affiliation{Instituto de F\'{i}sica Gleb Wataghin, Unicamp, 13083-859, Campinas-SP, Brazil}
\author{H. Freire}
\email{hermann\_freire@ufg.br}
\affiliation{Instituto de F\'{i}sica, Universidade Federal de Goi\'{a}s, 74.001-970,
Goi\^{a}nia-GO, Brazil}
\author{E. Miranda}
\email{emiranda@ifi.unicamp.br}
\affiliation{Instituto de F\'{i}sica Gleb Wataghin, Unicamp, 13083-859, Campinas-SP, Brazil}
\author{R. G. Pereira}
\email{rpereira@iip.ufrn.br}
\affiliation{International Institute of Physics and Departamento de F\'isica Te\'orica
e Experimental, Universidade Federal do Rio Grande do Norte, Campus Universit\'ario, Lagoa Nova, 
Natal, RN, 59078-970, Brazil}
\date{\today}

\begin{abstract}
We investigate the edge magnetism and the spin transport properties of an SU(2)-symmetric Kitaev spin liquid (KSL) model put forward by Yao and Lee [\href{https://link.aps.org/doi/10.1103/PhysRevLett.107.087205}{Phys. Rev. Lett. \textbf{107}, 087205 (2011)}] on the honeycomb lattice. In this model, the spin degrees of freedom fractionalize into a $\mathbb{Z}_{2}$ static gauge field and three species of either gapless (Dirac) or gapped (chiral) Majorana fermionic excitations. We find that, when a magnetic field is applied to a zigzag edge, the Dirac KSL exhibits a nonlocal magnetization associated with the existence of zero-energy edge modes. The application of a spin bias $V=\mu_{\uparrow}-\mu_{\downarrow}$ at the interface of the spin system with a normal metal produces a spin current into the KSL, which depends, in the zero-temperature limit, as a power-law on $V$ for both Dirac and chiral KSLs, but with different exponents. Lastly, we study the longitudinal spin Seebeck effect, in which a spin current is driven by  the combined action of a magnetic field perpendicular to the plane of the honeycomb lattice and a thermal gradient at the interface of the KSL with a metal. Our results suggest that edge magnetization and spin transport can be used to probe the existence of charge-neutral edge states in quantum spin liquids. 
\end{abstract}


\maketitle

\section{Introduction}

Spin fractionalization has become a key concept in the description of the physical properties of  quantum spin liquids (QSLs) \cite{Anderson-MRB(1973)}. The latter are highly entangled states of interacting magnetic moments, which break no internal symmetry, even at zero temperature, and can also support nonlocal anyonic excitations \cite{Balents-N(2010),Savary-RPP(2017)}. One example of such systems, known as the Kitaev spin liquid (KSL), was originally proposed as the exact ground state of a compass model on the honeycomb lattice \cite{Kitaev-AP(2006)}. In general, the KSL describes exchange-frustrated spin-$1/2$ systems defined on tricoordinated lattices \cite{Trebst-PRB(2016),Trebst-ArXiv(2017)}, in which the magnetic moments fractionalize into itinerant Majorana fermions and a static $\mathbb{Z}_{2}$ gauge field. Depending on the projective symmetries of the lattice, the Majorana fermion excitations can mimic the behavior of an electronic system harboring different nodal structures, namely  Dirac or Weyl points, nodal lines, and  Fermi surfaces. In addition, they exhibit Majorana surface states associated with the nontrivial momentum-space topology \cite{Trebst-PRB(2016),Surendran-PRB(2009)}. As demonstrated  by Jackeli and Khaliullin \cite{Jackeli-PRL(2009)}, the types of interactions present in a KSL model can be found naturally in Mott insulators of the $4d^{5}$ and $5d^{5}$ electronic configurations. Indeed, since this original proposal was put forward, several two- (2D) and three-dimensional (3D) compounds were identified as potential candidates for realizing the physics of KSLs \cite{Singh-PRB(2010),Williams-PRB(2016),Plumb-PRB(2014)}. 

To date, most theoretical predictions for   KSL materials have been addressed mainly by spectroscopic probes, including inelastic neutron scattering (INS) \cite{Banerjee-NM(2016),Banerjee-S(2017)}, Raman scattering \cite{Sandilands-PRL(2015),Sandilands-PRB(2016),Glamazda-NC(2016)}, and resonant inelastic x-ray scattering (RIXS) \cite{Gretarsson-PRB(2013)}. Although most of the  materials investigated exhibit some sort of magnetic order below a N\'{e}el temperature $T_{N}$ (with the notable exception of  H$_3$LiIr$_2$O$_6$ \cite{Kitagawa-Nature(2018)}), the experimental data reveal the presence of a continuum of fractional excitations at energy scales well above $T_{N}$,   consistent with the emergence of Majorana fermion excitations. More recently, the itinerant character of the excitations in the 2D KSL candidate $\alpha$-RuCl$_{3}$ has also been probed by measurements of the thermal conductivity \cite{Hirobe-PRB(2017),Leahy-PRL(2017)}. Remarkably,   quantum Monte Carlo calculations of the thermal conductivity for a pure Kitaev model on the honeycomb lattice have found that some features of the experimental data for that material can, in fact, be related to the emergence of a KSL state \cite{Nasu-PRL(2017)}.  

Another interesting avenue for investigating the properties of QSLs consists in the measurement of spin currents carried by the fractionalized  excitations. In this respect, Refs.\cite{Chen-PRB(2013),Sachdev-PRB(2015)} proposed an experimental setup in which a spin current is driven into a magnetic insulator by a spin bias produced by the spin Hall effect at the interface with a normal metal. According to this theory, if the  elementary spin excitations are treated as free quasiparticles  (spinons) within a mean-field approximation,  their low-energy dispersion can be inferred  from  the relation between the spin current and the spin bias. As an alternative to the spin Hall effect, spin currents can also be generated by the longitudinal spin Seebeck (LSS) effect \cite{Uchida-N(2008),Uchida-APL(2010),Kikkawa-PRL(2013)}, which results from a combination of a magnetic field and a temperature gradient at the interface of a magnetic insulator with a metal. For instance, a recent experiment   \cite{Saitoh-NP(2017)} has shown that the gapless spinons described by   Tomonaga-Luttinger (TL) liquid theory \cite{Giamarchi-QPOD(2003)} in the one-dimensional (1D) QSL material Sr$_{2}$CuO$_{3}$ give rise to an anomalous LSS effect, in which the flow of the spin current with respect to the magnetic field has an opposite sign when compared to that of an ordered magnetic insulator. In this system, the breaking  of  particle-hole symmetry of the   spinon spectrum is crucial to produce  a nonzero LSS response \cite{Saitoh-NP(2017)}.

\begin{figure}[t]
\centering
\includegraphics[width=1.0\linewidth]{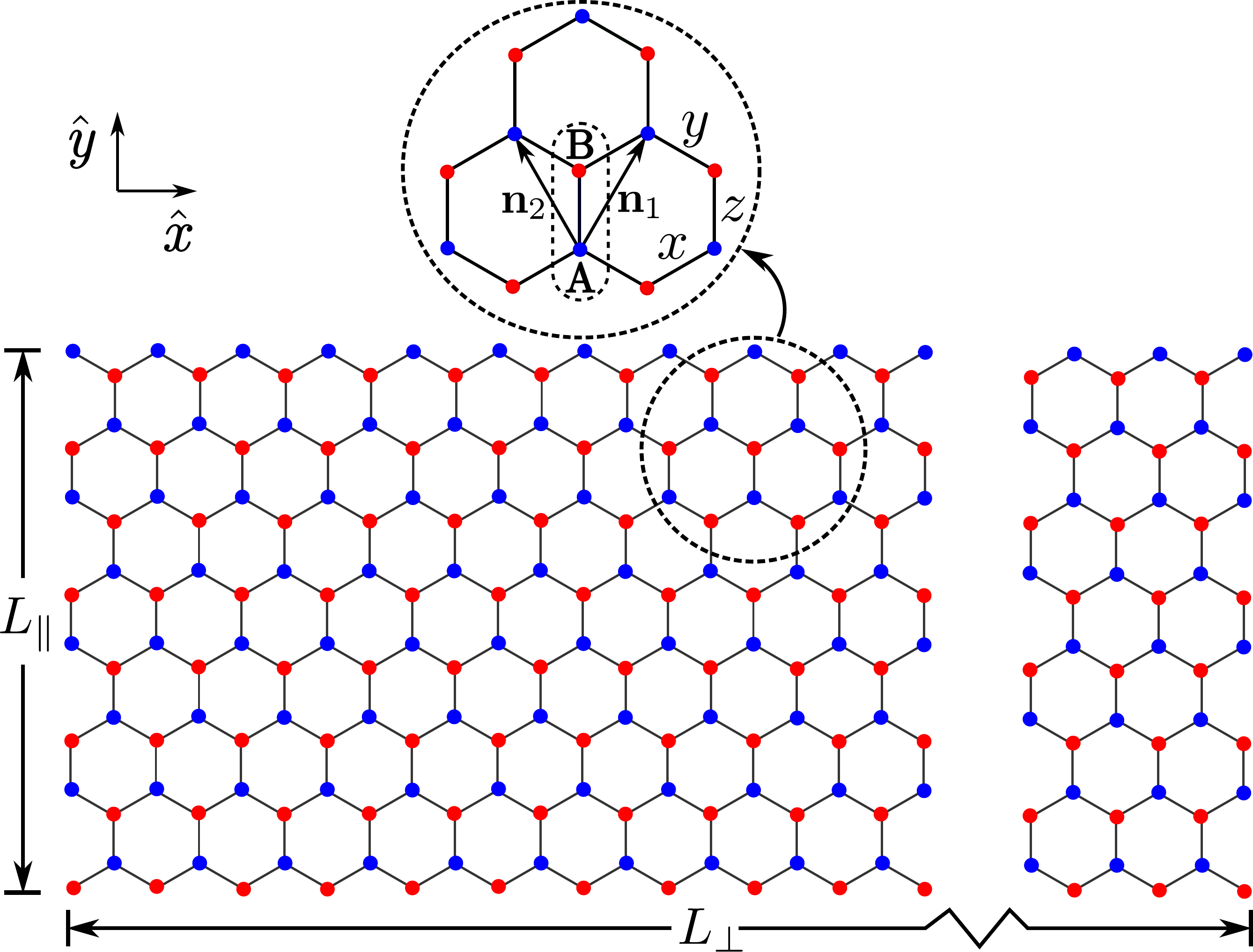}
\caption{(Color online) Schematic representation of the finite honeycomb lattice containing the degrees of freedom of the Yao-Lee model. Throughout this work, we consider a slab geometry with zigzag edges, such that the   lattice  obeys periodic boundary conditions   in the $\hat{\bm{x}}$ direction, and $L_{\perp}$  and $L_{\parallel}$ are defined as the unit-cell dimensions along the $\hat{\bm{x}}$ and $\hat{\bm{y}}$ axes, respectively. As depicted in the inset, the upper edge contains only sites of the A sublattice (blue circles, ``{\color{blue}$\bullet$}"), whereas the lower edge has sites of the B  sublattice (red circles, ``{\color{red}$\bullet$}"). The two vectors $\mathbf{n}_{1}=\frac{\sqrt{3}a}{2}(1,\sqrt{3})$, $\mathbf{n}_{2}=\frac{\sqrt{3}a}{2}(-1,\sqrt{3})$ refer to the lattice vectors. The allowed nearest-neighbor spin-exchange interactions $J_{x,y,z}$  are indicated   by the letters $x$, $y$, and $z$.}\label{Honeycomb_Lattice}
\end{figure}

Strictly speaking, in order for a spin current to propagate into a  system, the Hamiltonian must possess at least a global U(1) spin rotational symmetry. By contrast, in models with anisotropic spin interactions like  the original Kitaev model \cite{Kitaev-AP(2006)}, the total spin is not conserved and spin transport becomes ill-defined. However, over the past few years, several SU(2)-symmetric spin-$1/2$ models harboring KSL-like ground states have been proposed in different 2D lattices \cite{Wang-PRB(2010),Lee-PRL(2011),Sachdev-PRB(2011),Motrunich-PRB(2011)}. They share the common feature of being described by three degenerate species of free Majorana fermions hopping in the background of a $\mathbb{Z}_{2}$ gauge field. One example is the spin-$1/2$ Yao-Lee model \cite{Lee-PRL(2011)} defined on the honeycomb lattice. In the presence of time-reversal ($\mathcal{T}$) symmetry and spatially isotropic exchange interactions, its low-energy excitations are spin-$1$ gapless Majorana fermions with a linear dispersion around the Dirac point. When the $\mathcal{T}$ symmetry is broken by a next-nearest neighbor chiral interaction $J_{\chi}$, the Dirac points become gapped and the Majorana excitations behave like the spin-$1/2$ quasiparticles of a topological superconductor described by the SU(2)$_{2}$ Chern-Simons gauge theory. As a result, the system exhibits the spin quantum Hall effect with quantized spin Hall conductivity $\sigma^{s}_{xy}=1/(2\pi)$. 

In this work, we examine the edge magnetism and spin transport in the Yao-Lee model for both $\mathcal{T}$-invariant (Dirac) and $\mathcal{T}$-broken (chiral) KSL states. The exact solvability of the model obviates  the need for uncontrolled mean-field approximations in the description of the fractionalized quasiparticles. In order to calculate   static and dynamic response functions at the edge of the KSL, we numerically diagonalize the Yao-Lee Hamiltonian in a slab geometry with zigzag edges. This geometry is obtained by considering a honeycomb lattice with periodic boundary conditions along the $\hat{\bm{x}}$ direction and open boundary conditions for the $\hat{\bm{y}}$ direction, as schematically depicted in Fig. \ref{Honeycomb_Lattice}. First, we address the existence of the nonlocal effects in the KSL  by applying a local magnetic field at one edge of the system and then calculating the net magnetization at the opposite edge. We find that only in the case of the Dirac KSL, which has gapless excitations in the bulk and zero-energy flat bands as edge states, is  the opposite-edge magnetization finite for a large number of unit cells $L_{\parallel}$ separating the two edges.

Next, we calculate the spin current carried by the Majorana excitations into the KSL when a spin bias $V=\mu_{\uparrow}-\mu_{\downarrow}$ is applied at the KSL-metal interface. We show that the edge states give the dominant contribution to the net spin current when $|V|$ is much smaller than the Kitaev exchange interaction $J_{K}$. In the zero-temperature limit, this current depends as a power law $I_{\text{spin}}[V]\propto V^\alpha$ for both the Dirac and chiral KSL states. However, the exponents  are distinct and found here to be $\alpha_{\text{Dirac}}\approx 1$ and $\alpha_{\text{Chiral}}\approx 3$. The behavior of the spin current for the chiral KSL is equivalent to that for a 1D spinon Fermi sea \cite{Chen-PRB(2013)} and can be associated with the gapless chiral edge mode. Therefore, the result $I_{\text{spin}}[V]\propto V^3$ for the gapped chiral KSL is a consequence of its nontrivial topological properties.

Finally, we calculate the spin current driven by the LSS effect for a small temperature gradient between the metal and the KSL. Despite the intrinsic particle-hole symmetry of Majorana fermion excitations, we find a nonzero spin current, which is  again dominated by the contribution from the edge states. In addition, the spin current  changes sign as a function of  the chiral interaction $J_{\chi}$. In this respect, the KSL system is entirely distinct from the TL liquid, in which the unusual spin current due to the LSS effect is explained in terms of the breaking of the particle-hole symmetry in the spinon spectrum.  

This paper is structured as follows. In Section \ref{Section_S0}, we briefly review the Yao-Lee model and describe its exact solution using a Majorana representation for both   spin and orbital degrees of freedom. In Section \ref{Section_SI}, we show that a nonlocal edge magnetization occurs in this model in the absence of the $\mathcal{T}$-breaking interaction, which corresponds to the Dirac KSL state with zero-energy edge states. Section \ref{Section_SII} contains the calculation of the spin current in the KSL driven by a spin bias at the edge. In Section \ref{Section_SIII}, we show that the spin current can also be driven by the LSS effect, in which case it is controlled by a temperature gradient and an external magnetic field. Finally, Section \ref{Section_SIV} contains our concluding remarks. Details of some calculations are presented in the Appendix. 

\section{Model}\label{Section_S0}

The Yao-Lee model \cite{Lee-PRL(2011)} describes the interaction of spin-$1/2$ moments and orbital degrees of freedom residing on the sites of the honeycomb lattice (see Fig. \ref{Honeycomb_Lattice}). Its Hamiltonian is defined according to 
\begin{align}\label{Eq_YaoLee_Ham}
\mathcal{H}=&\ \sum\limits^{}_{\langle jk\rangle_{\alpha}}J_{\alpha}(\tau^{\alpha}_{j}\tau^{\alpha}_{k})\boldsymbol{\sigma}_{j}\cdot\boldsymbol{\sigma}_{k}\nonumber\\
&+J_{\chi}\sum\limits^{}_{\langle jk\rangle_{\alpha}}\sum\limits^{}_{\langle kl\rangle_{\beta}}\epsilon^{\alpha\beta\gamma}(\tau^{\alpha}_{j}\tau^{\gamma}_{k}\tau^{\beta}_{l})\boldsymbol{\sigma}_{j}\cdot\boldsymbol{\sigma}_{l},
\end{align}
where $J_{\alpha}$ ($\alpha\in\{x,y,z\}$) and $J_{\chi}$ are, respectively, the nearest-neighbor and the chiral next-nearest neighbor interaction. Here, $\boldsymbol{\sigma}_{j}$ and $\boldsymbol{\tau}_{j}$ denote two distinct vectors of Pauli matrices, which act, respectively, on the Hilbert space for spin and orbital degrees of freedom. At zero magnetic field, the model possesses a global SU(2) symmetry for the $\boldsymbol{\sigma}_{j}$ operators, but the interactions depend on the bond directions via the $\tau^{\alpha}_{j}$ operators. Although the Hamiltonian in Eq. \eqref{Eq_YaoLee_Ham} may  seem artificial at first, we should point out that similar models featuring   SU(2) symmetry and bond-directional interactions appear in Mott insulators with strong spin-orbit coupling and $j_{\text{eff}}=3/2$  local moments  \cite{Balents-PRB(2011),Pereira-PRL(2016),Jackeli-PRL(2017),Jackeli-ArXiv(2017)}.

\begin{figure}[t]
\centering
\subfloat[]{\includegraphics[width=0.92\linewidth]{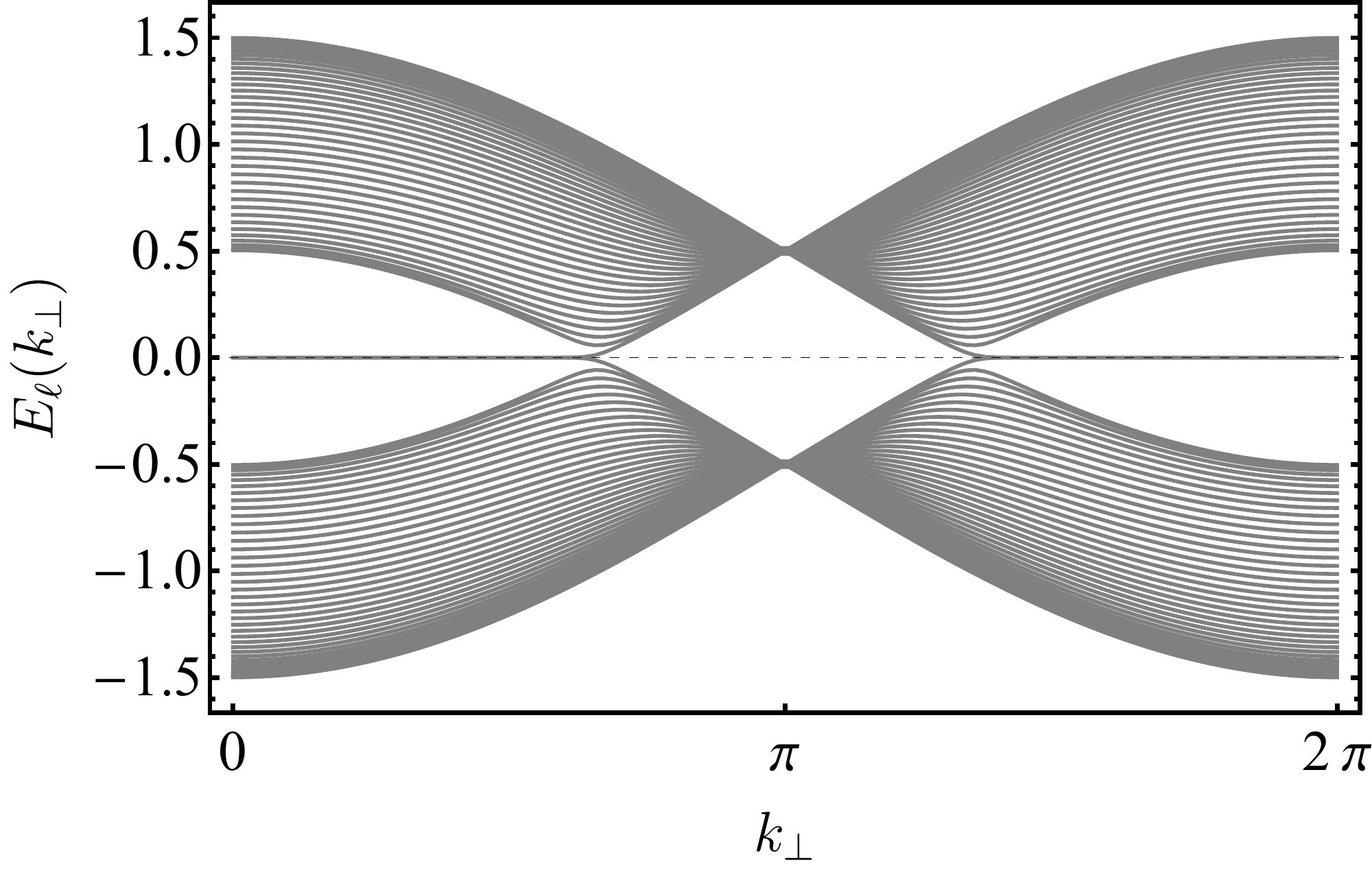}}
\vfill
\subfloat[]{\includegraphics[width=0.92\linewidth]{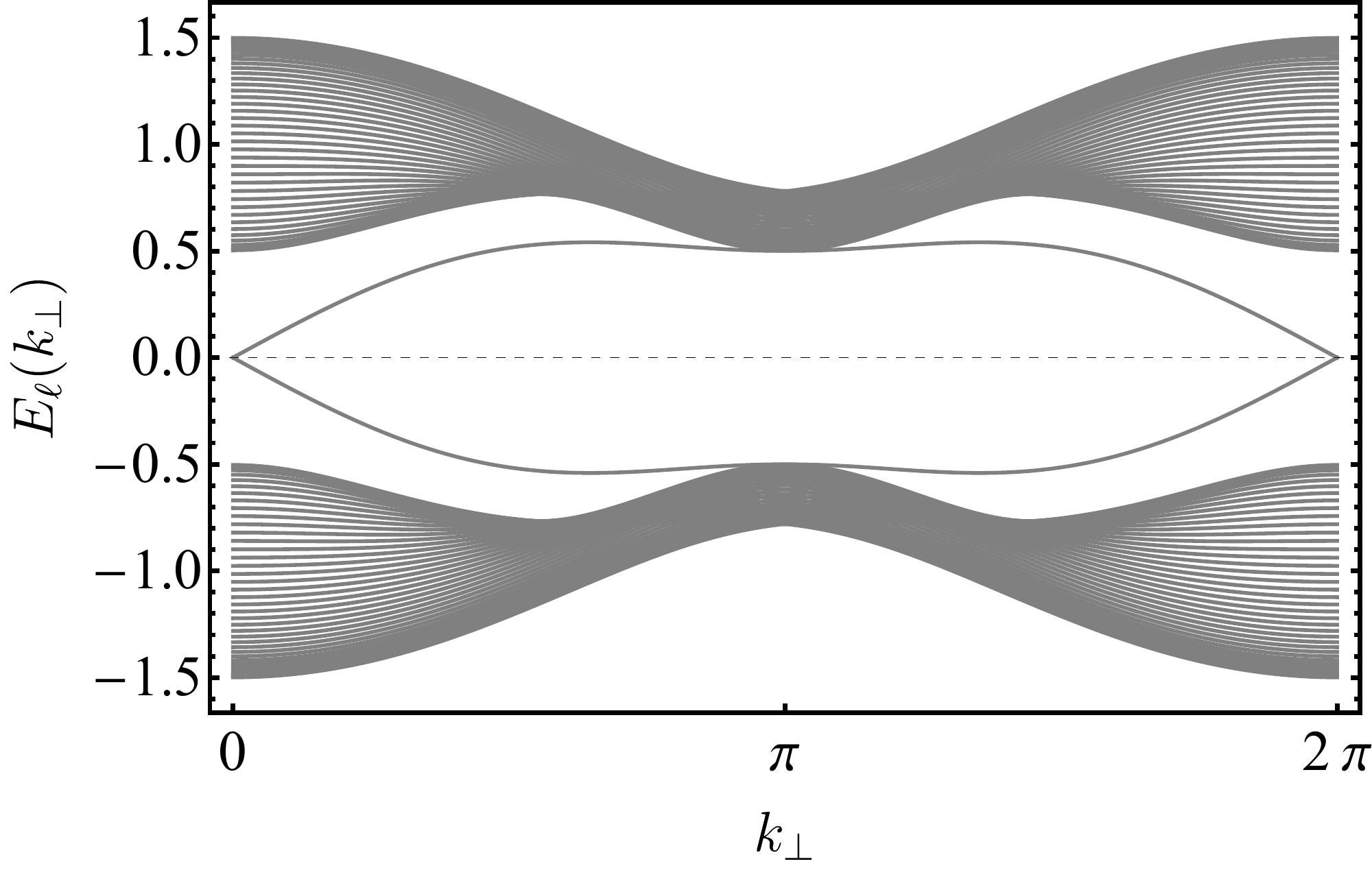}}
\caption{(Color online) Energy dispersion curves $E_{\ell}(k_{\perp})$ (in units of $J_K$) as a function of the $k_{\perp}$ momentum for the KSL on the honeycomb lattice with zigzag surfaces along the $\hat{\bm{y}}$ direction: (a) in the absence of the chiral interaction (\textit{i.e.}, $J_{\chi}=0$) and (b) for a finite chiral interaction (\textit{i.e.}, $J_{\chi}=0.3J_K$). Here, the lattice constant $a$ has been set to unity.}\label{Edge_States_SI}
\end{figure}

The Hamiltonian in Eq. \eqref{Eq_YaoLee_Ham} can be solved exactly by invoking a Majorana fermion representation for the spin-$1/2$ and the orbital operators \cite{Lee-PRL(2011),Sachdev-PRB(2011),Motrunich-PRB(2011),Tsvelik-PRL(1992),Coleman-PRB(1994)}. In order to do that, we write $\tau^{\alpha}_{j}$ and $\sigma^{\alpha}_{j}$ as
\begin{align}
\sigma^{\alpha}_{j}&=-\frac{i}{2}\epsilon^{\alpha\beta\gamma}c^{\beta}_{j}c^{\gamma}_{j},\label{Eq_SigmaRep}\\
\tau^{\alpha}_{j}&=-\frac{i}{2}\epsilon^{\alpha\beta\gamma}d^{\beta}_{j}d^{\gamma}_{j}\label{Eq_TauRep},
\end{align}
where $\eta^{\alpha}_{j}\in\{c^{\alpha}_{j},d^{\alpha}_{j}\}$ are Majorana fermion operators which, by definition, obey $\eta^{\alpha\dagger}_{j}=\eta^{\alpha}_{j}$ and the anti-commutation relations $\{\eta^{\alpha}_{j},\eta^{\beta}_{k}\}=2\delta_{\alpha\beta}\delta_{jk}$. In addition, the above transformations must be supplemented with the definition of the physical Hilbert space in terms of the operators $\eta_{j}$. This is achieved by demanding that the action of the projection operator $\hat{D}_{j}\equiv-ic^{x}_{j}c^{y}_{j}c^{z}_{j}d^{x}_{j}d^{y}_{j}d^{z}_{j}$ on every physical ket $|\Psi\rangle_{\text{phys}}$ is such that $\hat{D}_{j}|\Psi\rangle_{\text{phys}}=|\Psi\rangle_{\text{phys}}$. Restricting to spatially isotropic nearest-neighbor exchange interactions $J_{x}=J_{y}=J_{z}=J_{K}$ and   inserting the relations   in Eqs. \eqref{Eq_SigmaRep} and \eqref{Eq_TauRep} into Eq. \eqref{Eq_YaoLee_Ham}, we obtain
\begin{align}\label{Eq_MajHam}
\mathcal{H}=&\; J_{K}\sum\limits^{}_{\langle jk\rangle_{\alpha}}\sum\limits^{}_{\gamma=x,y,z}\hat{u}_{\langle jk\rangle_{\alpha}}ic^{\gamma}_{j}c^{\gamma}_{k}\nonumber\\
&+J_{\chi}\sum\limits^{}_{\langle jk\rangle_{\alpha}}\sum\limits^{}_{\langle kl\rangle_{\beta}}\sum\limits^{}_{\gamma=x,y,z}\hat{u}_{\langle jk\rangle_{\alpha}}\hat{u}_{\langle kl\rangle_{\beta}}ic^{\gamma}_{j}c^{\gamma}_{l},
\end{align}
where $\hat{u}_{\langle jk\rangle_{\alpha}}\equiv-id^{\alpha}_{j}d^{\alpha}_{k}$ are bond operators defined for two neighboring sites $j$ and $k$. Since $[\hat{u}_{\langle jk\rangle_{\alpha}},\hat{u}_{\langle lm\rangle_{\beta}}]=0$ and $[\hat{u}_{\langle jk\rangle_{\alpha}},\mathcal{H}]=0$, the operators $\hat{u}_{\langle jk\rangle_{\alpha}}$ are conserved and can be substituted by their eigenvalues $u_{\langle jk\rangle_{\alpha}}=\pm1$. Since the ground state of the present model is in the zero flux sector, the bond variables are chosen conventionally as $u_{\langle jk\rangle_{\alpha}}=1\;(-1)$ for a site $j=A\;(B)$. As a consequence, the spin system is described by a tight-binding Hamiltonian where three species of Majorana fermions $c^{\alpha}_{j}$ are coupled to  a static $\mathbb{Z}_{2}$ gauge field $u_{\langle jk\rangle_{\alpha}}$. Due to SU(2) spin symmetry of the Yao-Lee model, we also notice that the Hamiltonian in Eq. \eqref{Eq_MajHam} has an SO(3)$\otimes\mathbb{Z}_{2}$ invariance with respect to rotations of the three species of Majorana fermions and the gauge symmetry of the bond variables.

In Fig. \ref{Edge_States_SI}, we show the behavior of the energy dispersion curves $E_{\ell}(k_{\perp})$ of the Yao-Lee Hamiltonian in a slab geometry with zigzag edges as a function of momentum $k_{\perp}$ for $J_{\chi}=0$ and $J_{\chi}=0.3J_{K}$, which correspond, respectively, to the Dirac and chiral KSL states. In the absence of the chiral interaction, the system exhibits zero-energy flat bands on the edges. As $J_{\chi}$ becomes finite, the zero-energy flat bands give rise to a pair of dispersing edge modes, which are associated with the non-trivial Chern number $\nu_{\text{Chern}}=1$ and, thus, constitute a hallmark of the chiral KSL phase.

\section{Nonlocal edge magnetization}\label{Section_SI}

We now turn to the question of whether the KSL states of the Yao-Lee model on the honeycomb lattice exhibit any nonlocal property. This study is inspired by recent works on Weyl semimetals \cite{Vishwanath-PRB(2011)} reporting the emergence of nonlocal effects in their dc voltage and transmission of electromagnetic waves \cite{Parameswaran-PRX(2014),Baum-PRX(2015)}, and in the Coulomb drag for sheets of graphene separated by a slab of Weyl semimetal \cite{Baum-PRB(2017)}. In these latter systems, the nonlocal effects are due to either the chiral anomaly \cite{Adler-PR(1969),*Bell-NCA(1969),Nielsen-PLB(1983)}, which is effective in the bulk, or the topological Fermi arcs on the surface.

As the Yao-Lee model contains only charge-neutral spinful excitations, possible nonlocal effects can be effectively probed by measuring its magnetic response   under local magnetic fields. We consider a magnetic field $\bm{B}=B_{z}\hat{\bm{z}}$ applied to the spin-$1/2$ moments localized on the upper zigzag edge depicted in Fig. \ref{Honeycomb_Lattice}. This boundary magnetic field can be regarded as being due to the interface of the KSL with a ferromagnet (FM).  Note that only the $c^x$ and $c^y$ Majorana fermions couple to the magnetic field. In order to simplify the calculation, we use the complex fermions operators $f^{z}_{j}=(c^{x}_{j}-ic^{y}_{j})/2$, and write the corresponding part of the Hamiltonian in Eq. \eqref{Eq_MajHam} as 
\begin{align}
\mathcal{H}_{\text{FM}}=&\; 2J_{K}\sum\limits^{}_{\langle jk\rangle_{\alpha}}u_{\langle jk\rangle_{\alpha}}if^{z\dagger}_{j}f^{z}_{k}+\text{H.c.}\nonumber\\
&+2J_{\chi}\sum\limits^{}_{\langle jk\rangle_{\alpha}}\sum\limits^{}_{\langle kl\rangle_{\beta}}u_{\langle jk\rangle_{\alpha}}u_{\langle kl\rangle_{\beta}}if^{z\dagger}_{j}f^{z}_{l}+\text{H.c.}\nonumber\\
&-B_{z}\sum\limits^{}_{j\in\partial\mathcal{A}}\Bigl(f^{z\dagger}_{j}f^{z}_{j}-\frac{1}{2}\Bigr),
\end{align}
where $\sigma^{z}_{j}=(f^{z\dagger}_{j}f^{z}_{j}-1/2)$ is the spin operator along $\hat{\bm{z}}$, and $\partial\mathcal{A}$ represents the boundary  sites of the spin system which are coupled to  the magnetic field. We perform a partial Fourier transform of the $f^{z}_{j}$ fermions for momentum $\mathbf{k}_{\perp}=k_{\perp}\hat{\bm{x}}$, and then define the spinor $F^{z\dagger}(k_{\perp})=[f^{z\dagger}(k_{\perp},1),f^{z\dagger}(k_{\perp},2),\ldots,f^{z\dagger}(k_{\perp},L_{\parallel})]$, where $f^{z\dagger}(k_{\perp},y)=[f^{z\dagger}_{A}(k_{\perp},y),f^{z\dagger}_{B}(k_{\perp},y)]$. As a result, we obtain
\begin{equation}\label{Eq_FieldSlabHam}
\mathcal{H}_{\text{FM}}=\sum\limits^{}_{k_{\perp}}F^{z\dagger}(k_{\perp})\mathcal{H}_{\text{FM}}(k_{\perp},B_{z})F^{z}(k_{\perp})+\frac{L_{\perp}B_{z}}{2},
\end{equation}
where
\begin{equation}
\mathcal{H}_{\text{FM}}(k_{\perp},B_{z})=2\mathcal{H}(k_{\perp})+
\begin{pmatrix}
-B_{z} & \mathbb{0}_{1\times(2L_{\parallel}-1)} \\
\mathbb{0}_{(2L_{\parallel}-1)\times1} & \mathbb{0}_{(2L_{\parallel}-1)\times(2L_{\parallel}-1)}
\end{pmatrix}.
\end{equation}
Here, $\mathcal{H}(k_{\perp})$ is a matrix of dimension $2L_{\parallel}\times 2L_{\parallel}$ representing the slab Hamiltonian, which is given in the Appendix, whereas $\mathbb{0}_{n\times m}$ corresponds to a matrix of order $n\times m$, where all entries are zero.

The diagonalization of $\mathcal{H}_{\text{FM}}$ proceeds straightforwardly. By defining the canonical transformation
\begin{equation}\label{Eq_CanTrans}
F^{z}_{\ell}(k_{\perp})=\sum\limits^{2L_{\parallel}}_{\ell'=1}U_{\ell,\ell'}(k_{\perp},B_{z})\Upsilon^{z}_{\ell'}(k_{\perp},B_{z}),
\end{equation}
where $\Upsilon^{z}_{\ell}(k_{\perp},B_{z})$ are complex fermion operators, and $U(k_{\perp},B_{z})$ is a unitary matrix obeying
\begin{align}\label{Eq_DiagMatrix}
&U^{\dagger}(k_{\perp},B_{z})\mathcal{H}_{\text{FM}}(k_{\perp},B_{z})U(k_{\perp},B_{z})\nonumber\\
&=\operatorname{diag}\{E_{1}(k_{\perp},B_{z}),\ldots,E_{2L_{\parallel}}(k_{\perp},B_{z})\},
\end{align}
we find, after substituting Eqs. \eqref{Eq_CanTrans} and \eqref{Eq_DiagMatrix} into Eq. \eqref{Eq_FieldSlabHam}, that
\begin{align}
\mathcal{H}_{\text{FM}}=&\sum\limits^{2L_{\parallel}}_{\ell=1}\sum\limits^{}_{k_{\perp}}E_{\ell}(k_{\perp},B_{z})\Upsilon^{z\dagger}_{\ell}(k_{\perp},B_{z})\Upsilon^{z}_{\ell}(k_{\perp},B_{z})\nonumber\\
&+L_{\perp}B_{z}/2.
\end{align} 

\begin{figure*}[t]
\centering
\subfloat[]{\includegraphics[width=0.495\linewidth]{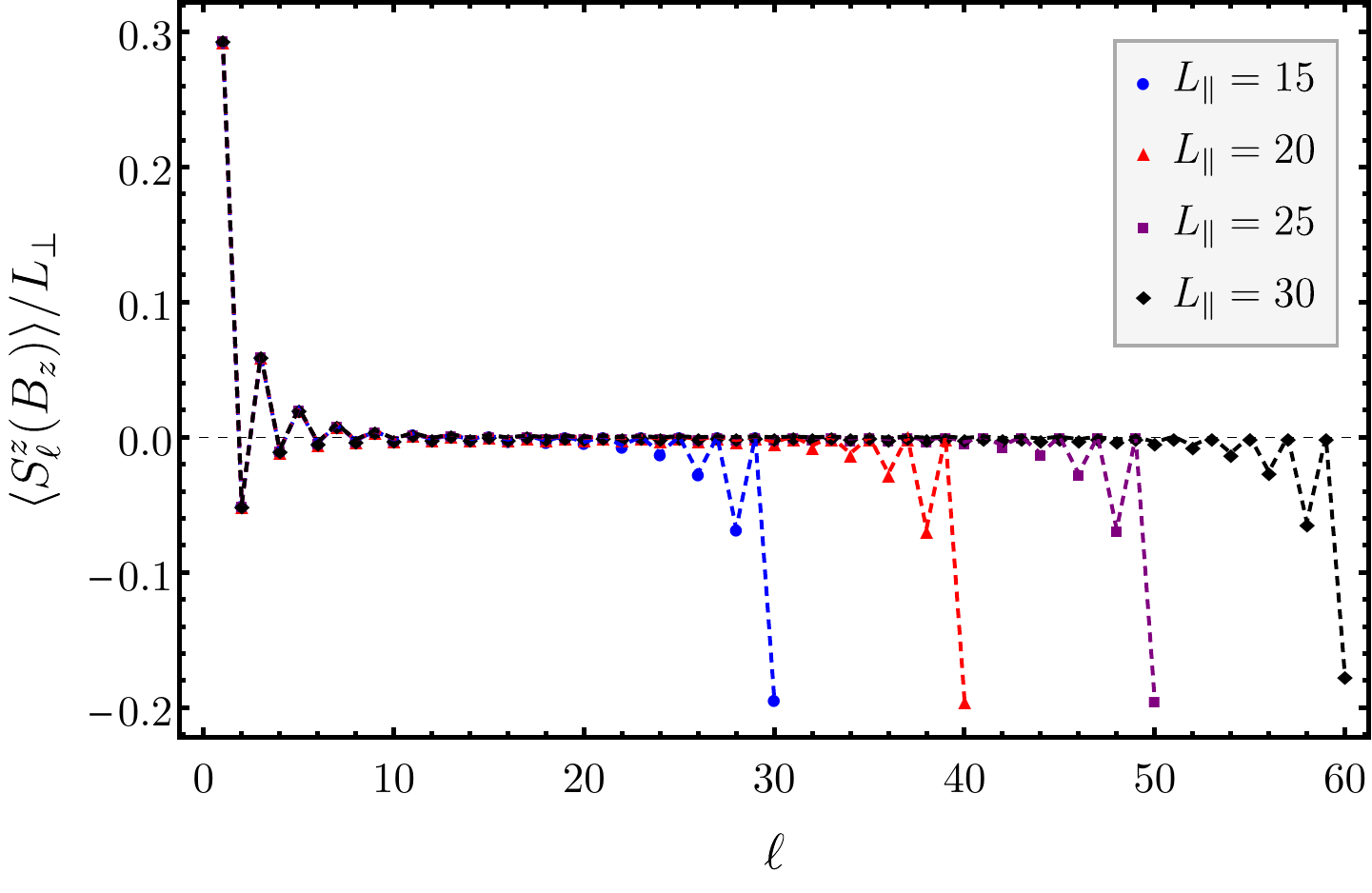}\label{Non_Local_01}}
\hfill
\subfloat[]{\includegraphics[width=0.490\linewidth]{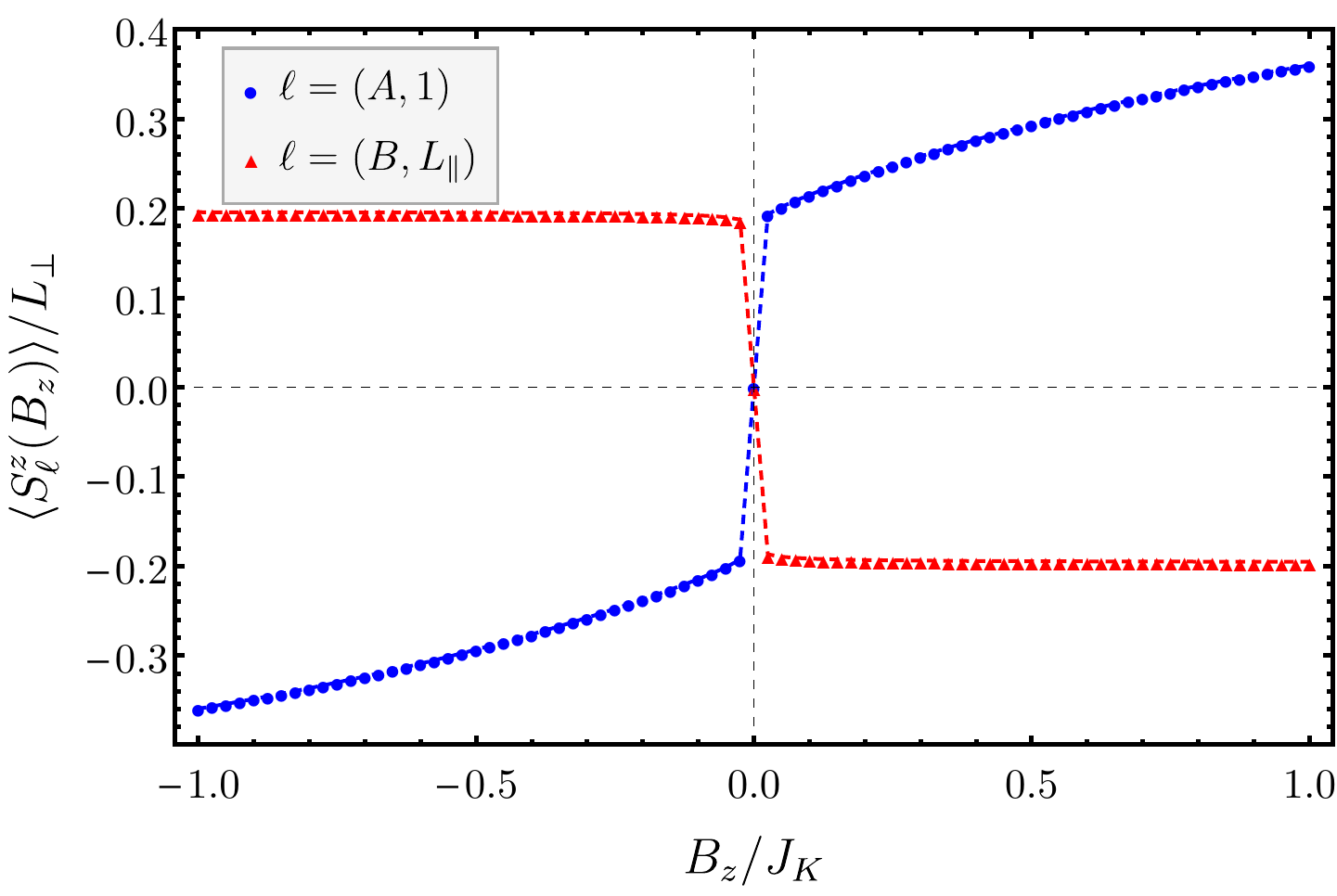}\label{Non_Local_02}}
\vfill
\subfloat[]{\includegraphics[width=0.490\linewidth]{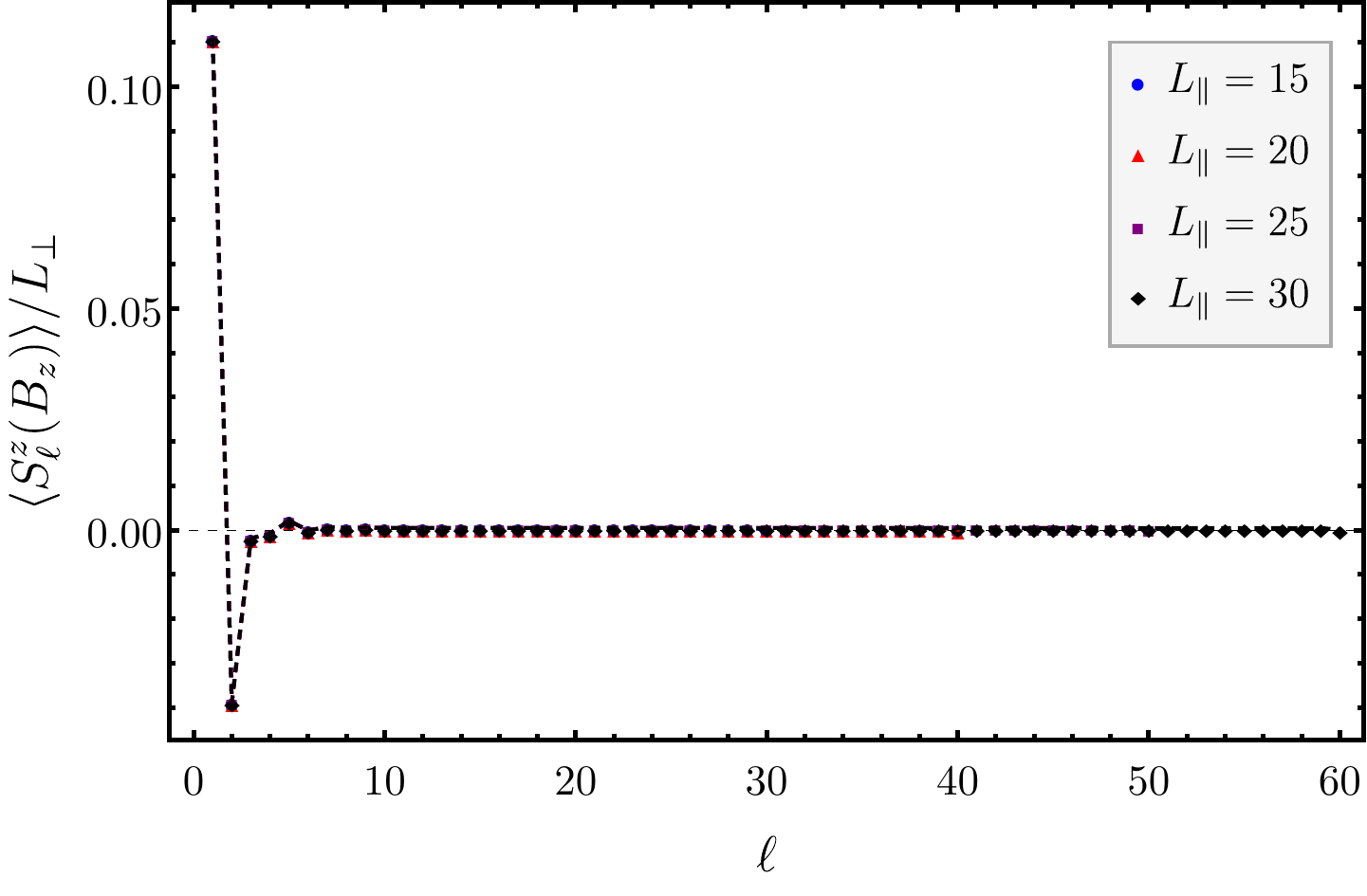}\label{Non_Local_03}}
\hfill
\subfloat[]{\includegraphics[width=0.490\linewidth]{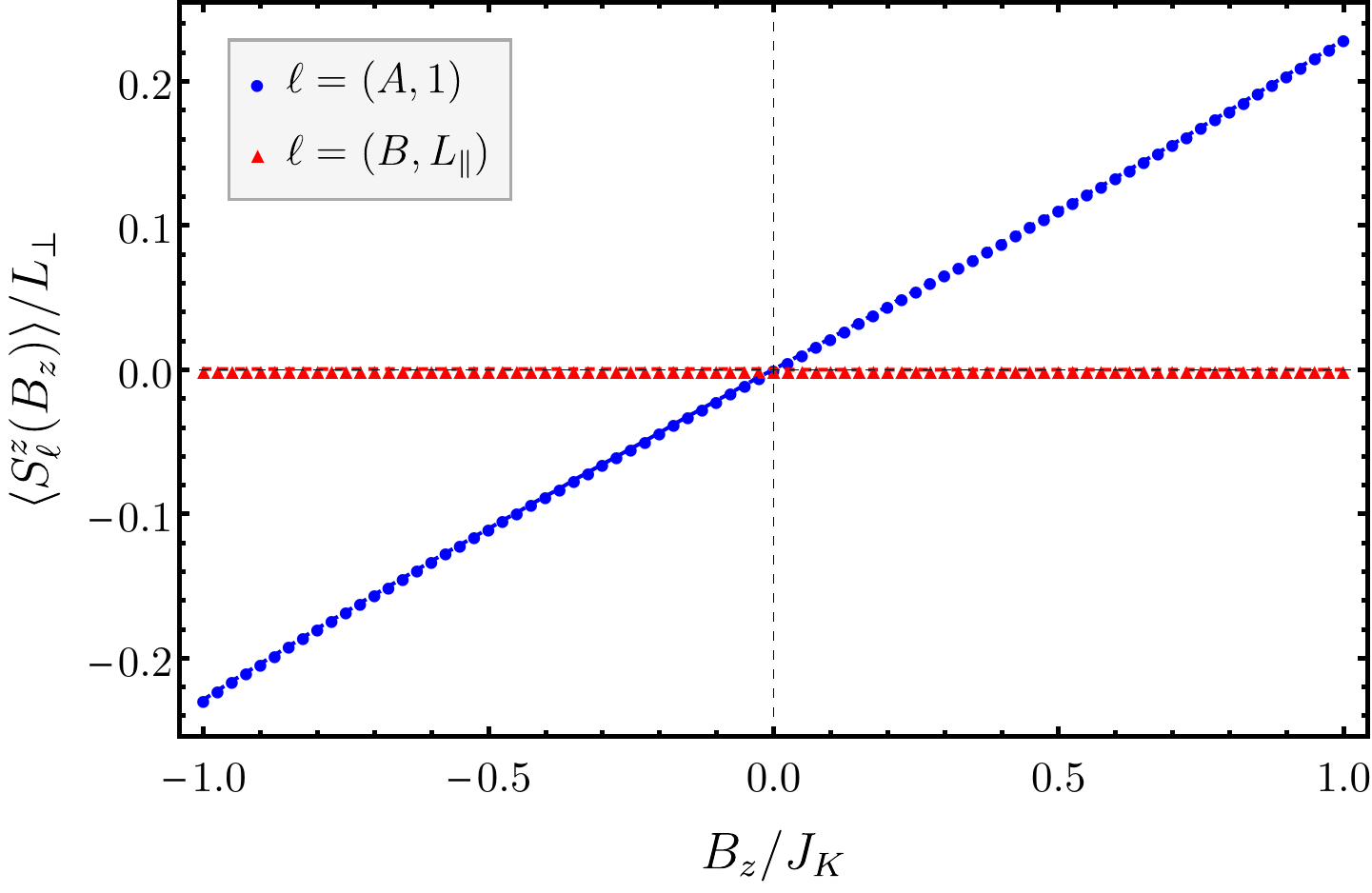}\label{Non_Local_04}}
\caption{(Color online) Magnetization per unit cell $\langle S^{z}_{\ell}(B_{z})\rangle/L_{\perp}$ in the limit of $T\rightarrow 0$ for the sublattices $\ell=(b,L_{\parallel}-y+1)$ parallel to the $\hat{\bm{x}}$ direction. This magnetization is induced by a local magnetic field acting on the spin-$1/2$ magnetic moments at the upper zigzag edge of the honeycomb lattice shown in Fig. \ref{Honeycomb_Lattice}. Panels (a) and (b) refer to the behavior exhibited by the Dirac KSL, whereas panels (c) and (d) refer to the behavior of the chiral KSL. In (a) and (c), we set $B_z=0.5J_K$. Note that a nonlocal (long-distance) edge magnetization appears just for the spin system described by the Dirac KSL. These results were obtained by substituting the momentum integral in Eq. \eqref{Eq_EdgeMag} by a sum over a mesh with $1500$ points.}
\end{figure*}

From the above result, we can finally calculate the zero-temperature magnetization per unit cell for the spins of a sublattice $b\in\{A,B\}$ parallel to $\hat{\bm{x}}$ (see Fig. \ref{Honeycomb_Lattice}), which is specified here by the composite index $\ell=(b,L_{\parallel}-y+1)$ \footnote{Here, we point out that the index $\ell$ denotes the position on the lattice. It should not be confused with the index $\ell$ in $E_{\ell}(k_{\perp})$ that is a label to represent the energy dispersion curves of the model.}. Indeed, one obtains
\begin{align}\label{Eq_EdgeMag}
\frac{\langle S^{z}_{\ell}(B_{z})\rangle}{L_{\perp}}\stackrel{T\rightarrow 0}{=}&\sum\limits^{2L_{\parallel}}_{\ell'=1}\int\frac{dk_{\perp}}{2\pi}|U_{\ell,\ell'}(k_{\perp},B_{z})|^{2}\Theta[-E_{\ell'}(k_{\perp},B_{z})]\nonumber\\
&-1/2,
\end{align}
where $\Theta(x)$ is the Heaviside step function.

In Figs. \ref{Non_Local_01} and \ref{Non_Local_02}, we show the behavior of $\langle S^{z}_{\ell}(B_{z})\rangle/L_{\perp}$ for the Dirac KSL ($J_{\chi}=0$). Figure \ref{Non_Local_01} refers to the dependence of $\langle S^{z}_{\ell}(B_{z})\rangle/L_{\perp}$ on the sublattice index $\ell$ for systems with different edge distances $L_{\parallel}$ and fixed magnetic field $B_{z}=0.5J_{K}$. Figure \ref{Non_Local_02} shows the magnetic field dependence of $\langle S^{z}_{\ell}(B_{z})\rangle/L_{\perp}$ on the two sublattices $\ell=(A,1)$ and $\ell=(B,L_{\parallel})$, which are localized at the two opposite edges for a system with $L_{\parallel}=25$. We observe that both the magnetization at the edge $\ell=(A,1)$, which is under the influence of the local magnetic field, and the one at the opposite edge $\ell=(B,L_{\parallel})$ are finite for non-zero magnetic fields, but point in different directions. Besides, the magnetization oscillates between positive and negative values with the sublattice index $\ell$. In Figs. \ref{Non_Local_03} and \ref{Non_Local_04}, we show the same results for the chiral KSL with chiral interaction given by $J_{\chi}=0.3J_{K}$. In this case, we find, however, that the system exhibits a finite magnetization only in the vicinity of the edge $\ell=(A,1)$ under the action of the local magnetic field. We have also checked that these results do not change qualitatively for other finite values of $J_{\chi}$.

The nonlocal magnetization for the Dirac KSL can be explained in terms of the zero-energy edge bands   and the long-range correlations among the spin-$1/2$ moments. Indeed, since the    edge states have no kinetic energy, they can order ferromagnetically by the action of an infinitesimal magnetic field. However, the magnetic field here is restricted to the vicinity of just one edge. Due to the power-law decaying spin correlations present in this state, which behave as $\langle\sigma^{z}(\mathbf{r})\sigma^{z}(0)\rangle\sim1/|\mathbf{r}|^{4}$ \cite{Lee-PRL(2011)} for spins separated by a distance $|\mathbf{r}|\gg 1$, the action of a magnetic field at one edge can create a small perturbation at the opposite edge and, therefore, polarize the zero-energy modes. Consequently, we expect that other SU(2)-symmetric Kitaev spin models with long-range correlations and flat surface bands, which can be realized in the so-called harmonic-honeycomb lattices \cite{Kim-PRL(2015)}, could also exhibit the same sort of nonlocal magnetization as described here.

\section{Spin currents carried by Majorana fermions}\label{Section_SII}

In this section, we are concerned with the evaluation of a spin current injected from a normal metal with a non-equilibrium distribution of spins into the KSL described by Eq. \eqref{Eq_YaoLee_Ham} and carried by the Majorana fermion excitations. This calculation is based on the general spin-transport theory proposed by Chatterjee and Sachdev \cite{Sachdev-PRB(2015)}, which can be applied to magnetic insulators with and without long-range order. 

\begin{figure}[t]
\centering
\includegraphics[width=0.95\linewidth]{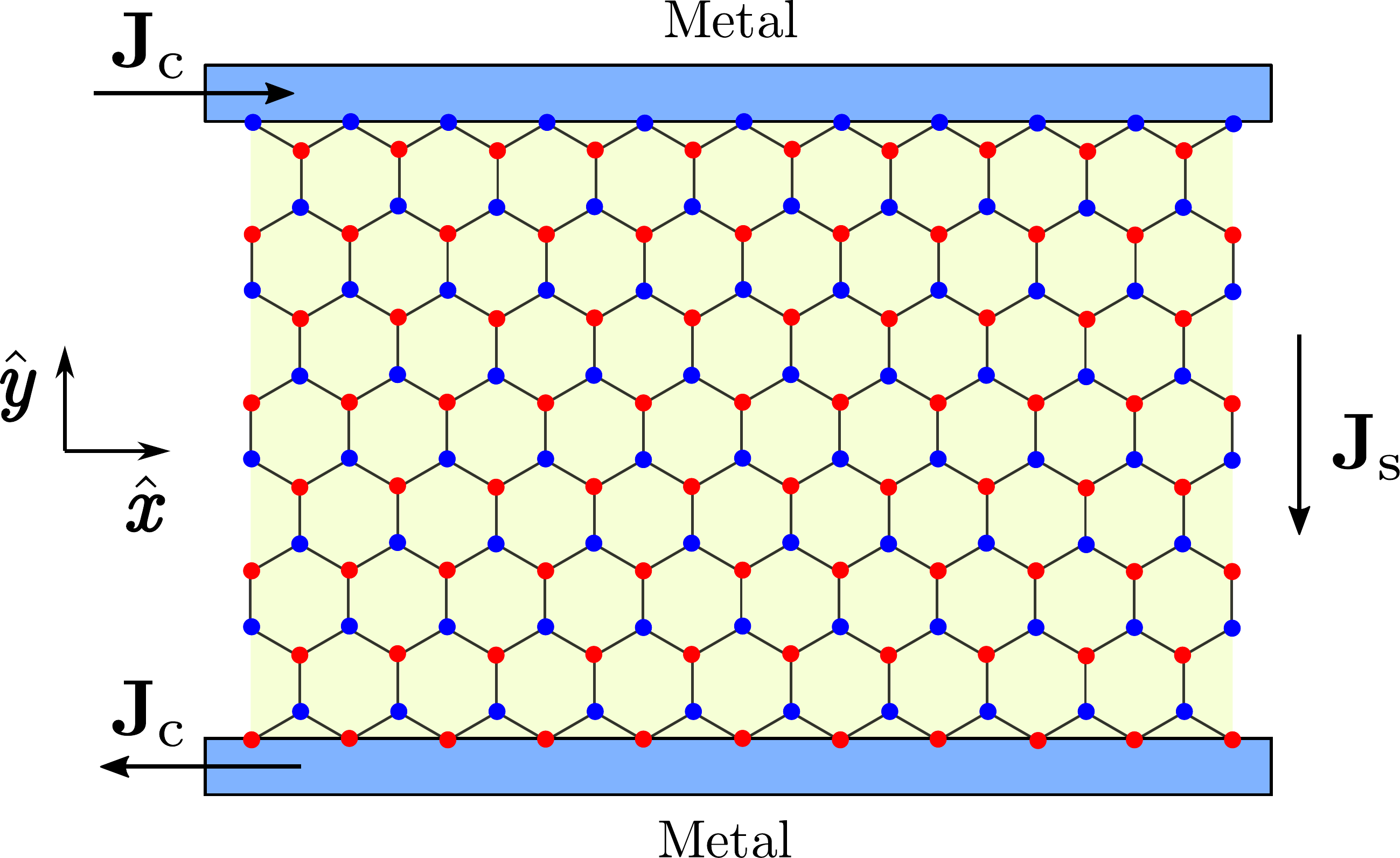}
\caption{(Color online) Schematic representation of the experimental setup used for the injection of spin currents into the KSL described by the Yao-Lee model. In the presence of strong spin-orbital coupling or disordered skew scattering, an electronic current density $\mathbf{J}_{\text{C}}$ flowing in a metallic plate produces a transverse spin current density $\mathbf{J}_{\text{S}}$ at the KSL-metal interface. As a result, the spin current is injected into the KSL and carried here by the Majorana fermionic excitations, originated from the fractionalization of the magnetic moments. This unusual spin current can be detected by measuring a finite charge current $\mathbf{J}_{\text{C}}$ produced in a second metallic plate by the inverse spin Hall effect.}\label{Sachdev_Setup}
\end{figure}

In the present case, the experimental setup for the generation of a nonzero spin current carried by the Majorana fermions consists of the KSL on the honeycomb lattice sandwiched between two metallic plates, as schematically depicted in Fig. \ref{Sachdev_Setup}. It is assumed that these metallic plates behave as a conventional Fermi liquid metal with quadratic band dispersion and Fermi energy $\epsilon_{F}$. According to this setup, a charge current $\mathbf{J}_{\text{C}}$ flowing in a metallic plate subjected to strong spin-orbit coupling or disordered skew scattering generates a transverse spin current density $\mathbf{J}_{\text{S}}$ through the edge boundary with the KSL. As a consequence, a spin current is injected into the KSL and carried here by the Majorana fermion excitations to the opposite edge in contact with a second metallic plate, where it can be detected by measuring the charge current $\mathbf{J}_{\text{C}}$ produced by the inverse spin Hall effect. In the case where the KSL-metal system is in thermal equilibrium, the spin current is basically a function of the spin bias (or spin chemical potential difference) $V=\mu_{\uparrow}-\mu_{\downarrow}$ between spin-up and spin-down electrons, which is controlled by the charge current at the first metallic plate. In fact, according to the theory proposed in Ref. \cite{Sachdev-PRB(2015)}, the spin current $I_{\text{spin}}$ flowing into the KSL is given by
\begin{equation}
I_{\text{spin}}[V]=I_{\text{spin},\uparrow}[V]-I_{\text{spin},\downarrow}[V],
\end{equation}
where
\begin{align}
I_{\text{spin},\uparrow}[V]=&\frac{\pi J^{2}L_{\perp}\nu(\epsilon_{F})}{2}\int^{V}_{0}\frac{d\omega}{2\pi}\int\frac{dq_{\perp}}{2\pi}\frac{V-\omega}{1-e^{-\beta(V-\omega)}}\nonumber\\
&\times S_{-+}(\bm{q}_{\perp},\omega),
\end{align}
\begin{align}
I_{\text{spin},\downarrow}[V]=&\frac{\pi J^{2}L_{\perp}\nu(\epsilon_{F})}{2}\int^{V}_{0}\frac{d\omega}{2\pi}\int\frac{dq_{\perp}}{2\pi}\frac{V+\omega}{e^{\beta(V+\omega)}-1}\nonumber\\
&\times S_{+-}(\bm{q}_{\perp},\omega).\label{Eq_Down_I}
\end{align}
Here, $J$ is the coupling of the metal with the spin system, $L_{\perp}$ is the interface area (length) perpendicular to the spin current, $\nu(\epsilon_{F})$ is the density of states at the Fermi energy, and $\beta=1/T$ is the reciprocal temperature. $I_{\text{spin},\uparrow}[V]$ describes the spin current due to spin-up electrons flipping to spin-down ones, whereas $I_{\text{spin},\downarrow}[V]$ corresponds to the spin current resulting from spin-down electrons flipping to spin-up ones. To evaluate $I_{\text{spin}}[V]$ at finite temperature, one needs to determine the dynamical spin structure factors (DSSFs) $S_{-+}(\bm{q}_{\perp},\omega)$ and $S_{+-}(\bm{q}_{\perp},\omega)$ at the boundary of the spin system. For the former DSSF, one has the definition
\begin{align}\label{Eq_SachdevDSSF01}
&S_{-+}(\bm{q}_{\perp},\omega)\nonumber\\
&=\frac{1}{L_{\perp}}\sum\limits^{}_{l,j\in\partial\mathcal{A}}e^{-i\bm{q}_{\perp}\cdot(\bm{X}_{l}-\bm{X}_{j})}\int^{\infty}_{-\infty}dt\; e^{i\omega t}\langle\sigma^{-}_{l}(t)\sigma^{+}_{j}(0)\rangle,
\end{align} 
where the indices $l$ and $j$ refer here to   sites at the upper interface $\partial\mathcal{A}$ of the metal with the KSL shown in Fig. \ref{Sachdev_Setup} and $\sigma^{\pm}_{l}(\tau)\equiv e^{\tau\mathcal{H}}\sigma^{\pm}_{l}e^{-\tau\mathcal{H}}$ denotes the imaginary-time Heisenberg representation of the transverse spin operators $\sigma^{\pm}_{l}=\sigma^{x}_{l}\pm i\sigma^{y}_{l}$. 

In order to determine $S_{-+}(\bm{q}_{\perp},\omega)$, we first calculate the Matsubara spin correlation function
\begin{align}\label{Eq_Mats01}
&X_{-+}(\bm{q}_{\perp},i\omega_{n})\nonumber\\
&=\frac{1}{L_{\perp}}\sum\limits^{}_{l,j\in\partial\mathcal{A}}e^{-i\bm{q}_{\perp}\cdot(\bm{X}_{l}-\bm{X}_{j})}\int^{\beta}_{0}d\tau e^{i\omega_{n}\tau}\langle T_{\tau}[\sigma^{-}_{l}(\tau)\sigma^{+}_{j}(0)]\rangle,
\end{align}
where the wave vector $\bm{q}_{\perp}$ points along $\hat{\bm{x}}$, since we are considering a honeycomb lattice with periodic boundary conditions along this direction. To specify the position $\bm{X}_{l}$ of a site $l$ of the honeycomb lattice, we make the substitution $l\rightarrow(b,\bm{R})$, where $b\in\{A,B\}$ refers as before to the sublattice index, and $\bm{R}$ represents a lattice vector. Using this notation and setting $\bm{d}\equiv a\frac{\sqrt{3}}{2}\hat{\bm{y}}$, we obtain
\begin{equation}
\bm{X}_{(b,\bm{R})}=\bm{R}+\varphi(b)\bm{d},
\end{equation}
where $\varphi(b)=0\; (1)$ for $b=A\; (B)$. 

Next, we use the Majorana representation for the spin operators and then rewrite the spin correlation function $\langle T_{\tau}[\sigma^{-}_{l}(\tau)\sigma^{+}_{j}(0)]\rangle$ in terms of them. Since the the DSSFs are calculated in the absence of a magnetic field, the Hamiltonian in Eq. \eqref{Eq_MajHam} for this situation has no cross terms involving Majorana fermions of different species. As a result, $\langle T_{\tau}[\sigma^{-}_{l}(\tau)\sigma^{+}_{j}(0)]\rangle$ can be factorized, according to the Wick's theorem, into two-point correlation functions involving Majorana fermions of the same species. In addition, the SU(2) symmetry of the Yao-Lee Hamiltonian implies that the correlation functions for the three species of Majorana fermions are equal to each other. As a result, we obtain
\begin{equation}\label{Eq_Wick}
\langle T_{\tau}[\sigma^{-}_{b}(\bm{R},\tau)\sigma^{+}_{b'}(\bm{R}',0)]\rangle=\frac{1}{2}\langle T_{\tau}[c^{\gamma}_{b}(\bm{R},\tau)c^{\gamma}_{b'}(\bm{R}',0)]\rangle^{2},
\end{equation}
where the superscript $\gamma$ represents just one of the three Majorana fermion species. The correlation function in Eq. \eqref{Eq_Wick} can now be evaluated by performing the partial Fourier transform of the Majorana operators and then rewriting the resulting $k_{\perp}$-dependent Majorana operators according to the unitary transformation that diagonalizes the   Hamiltonian for the slab geometry in Fig. \ref{Honeycomb_Lattice}. After performing this calculation, we obtain that the correlation function in Eq. \eqref{Eq_Mats01} evaluates to
\begin{widetext}
\begin{align}\label{Eq_Mats02}
X_{-+}(q_{\perp},i\omega_{n})=&\ \frac{1}{2L_{\perp}}\sum\limits^{2L_{\parallel}}_{\ell_{1}=1}\sum\limits^{2L_{\parallel}}_{\ell_{2}=1}\sum\limits^{}_{k_{\perp}}\Bigl\vert\sum\limits^{}_{\ell\in\partial\mathcal{A}}U_{\ell,\ell_{1}}(k_{\perp})U_{\ell,\ell_{2}}(-k_{\perp}-q_{\perp})\Bigr\vert^{2}\mathit{n}_{F}[-E_{\ell_{1},z}(k_{\perp})]\mathit{n}_{F}[-E_{\ell_{2},z}(-k_{\perp}-q_{\perp})]\nonumber\\
&\times\frac{\exp\{-\beta[E_{\ell_{1},z}(k_{\perp})+E_{\ell_{2},z}(-k_{\perp}-q_{\perp})]\}-1}{i\omega_{n}-E_{\ell_{1},z}(k_{\perp})-E_{\ell_{2},z}(-k_{\perp}-q_{\perp})},
\end{align}
\end{widetext}
where $E_{\ell,z}(k_{\perp})=2E_{\ell}(k_{\perp})$ and $U_{\ell,\ell'}(k_{\perp})$ are, respectively, the double eigenvalues and eigenvectors of the slab Hamiltonian for zigzag surfaces (see Appendix), and $\mathit{n}_{F}(\epsilon)=(e^{\beta\epsilon}+1)^{-1}$ refers to the Fermi-Dirac distribution function. Finally, the DSSF can be obtained in the following way
\begin{align}\label{DSSF}
S_{-+}(q_{\perp},\omega)=\operatorname{Im}[X_{-+}(q_{\perp},i\omega_{n}\rightarrow\omega+i\eta)],
\end{align}
where $\eta\rightarrow 0^{+}$. 

The procedure for calculating $S_{+-}(q_{\perp},\omega)$ follows the same steps we used to evaluate $S_{-+}(q_{\perp},\omega)$. However, we will not determine here $S_{+-}(q_{\perp},\omega)$, because we are mostly interested in the zero-temperature limit of the spin current $I_{\text{spin}}[V]$ for spin bias $V=\mu_{\uparrow}-\mu_{\downarrow}>0$. In this limit, $I_{\text{spin},\downarrow}[V]$ vanishes according to Eq. (\ref{Eq_Down_I}). Due to this fact, the total spin current for $T\rightarrow 0$ evaluates to 
\begin{equation}
I_{\text{spin}}[V]=\hspace{-0.1cm}\frac{\pi J^{2}L_{\perp}\nu(\epsilon_{F})}{2}\hspace{-0.1cm}\int^{V}_{0}\hspace{-0.1cm}\frac{d\omega}{2\pi}(V-\omega)\hspace{-0.1cm}\int\frac{dq_{\perp}}{2\pi}S_{-+}(q_{\perp},\omega),
\end{equation}
where the zero-temperature limit of $S_{-+}(q_{\perp},\omega)$ for positive frequencies is given by
\begin{widetext} 
\begin{align}\label{Eq_SachdevDSSF02}
S_{-+}(q_{\perp},\omega)\stackrel{T\rightarrow 0}{=}&\ \frac{\pi}{2}\sum\limits^{2L_{\parallel}}_{\ell_{1}=1}\sum\limits^{2L_{\parallel}}_{\ell_{2}=1}\int\frac{dk_{\perp}}{2\pi}\Bigl\vert\sum\limits^{}_{\ell\in\partial\mathcal{A}}U_{\ell,\ell_{1}}(k_{\perp})U_{\ell,\ell_{2}}(-k_{\perp}-q_{\perp})\Bigr\vert^{2}\Theta[E_{\ell_{1},z}(k_{\perp})]\Theta[E_{\ell_{2},z}(-k_{\perp}-q_{\perp})]\nonumber\\
&\times\delta[\omega-E_{\ell_{1},z}(k_{\perp})-E_{\ell_{2},z}(-k_{\perp}-q_{\perp})].
\end{align}
\end{widetext}
We should mention here that the above expression is exact for $T\rightarrow 0$, because in this limit we do not have to include the fluctuations of the $\mathbb{Z}_{2}$ gauge field. This should be contrasted with the DSSFs for the Kitaev model, where spin operators involves the insertion of a flux pair \cite{Knolle-PRL(2014)}. Moreover, since $S_{-+}(q_{\perp},\omega)$ depends in this limit on two Heaviside step functions, whose arguments are the energy dispersions of the slab Hamiltonian, we have, therefore, to take into account only its positive energies states for the evaluation of this DSSF.

\begin{figure}[t]
\centering
\includegraphics[width=1.0\linewidth]{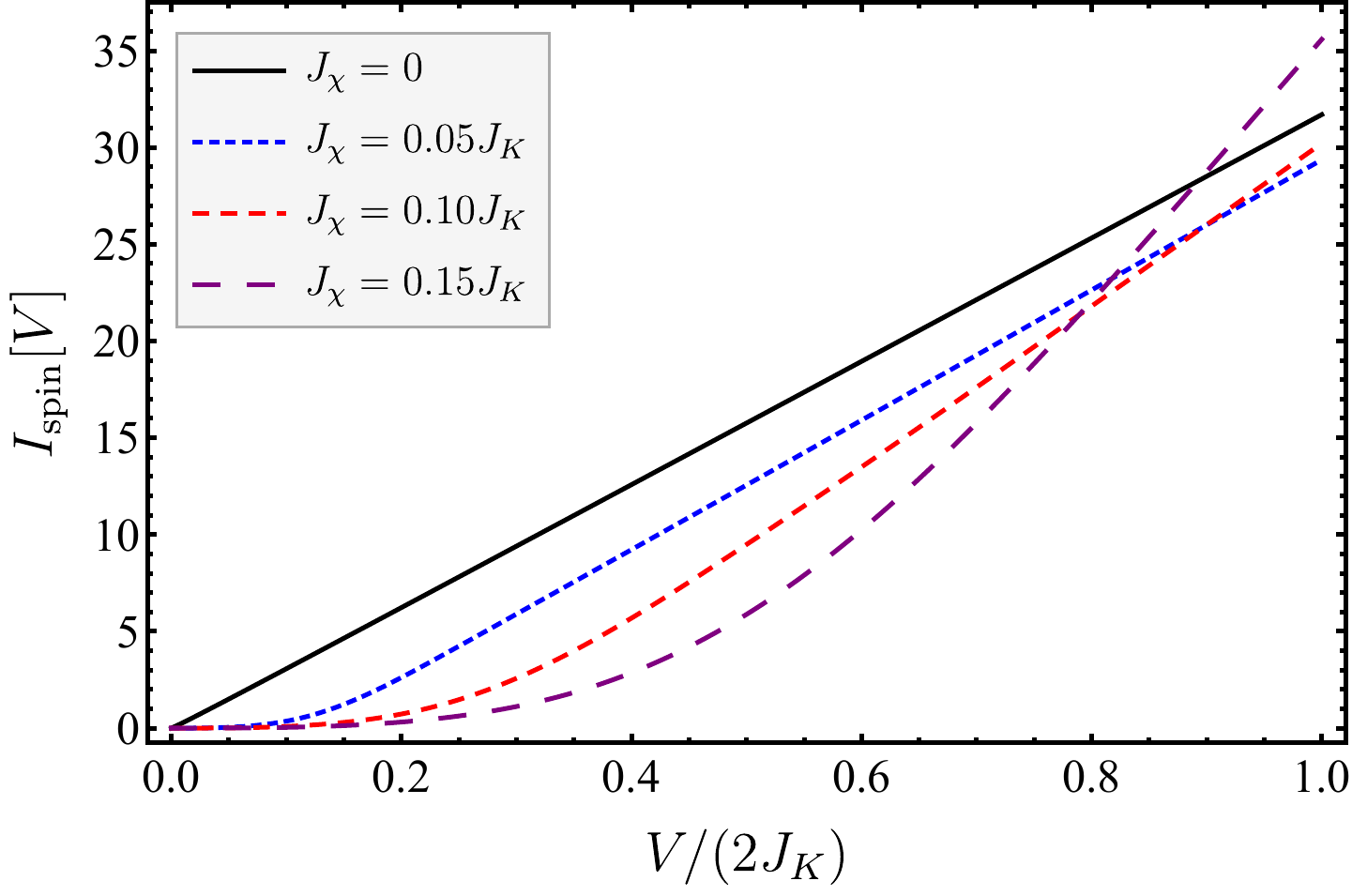}
\caption{(Color online) Zero-temperature limit of the spin current $I_{\text{spin}}[V]$ [in units of $\pi J^{2}L_{\perp}\nu(\epsilon_{F})J_{K}/4$] for chiral interactions $J_{\chi}=0,\;0.05J_{K},\;0.10J_{K},\;0.15J_{K}$. The spin current for the Dirac KSL behaves as a linear function of the spin bias $V$. On the other hand, a finite $J_{\chi}$ (\textit{i.e.}, chiral KSL) leads to the cubic dependence of $I_{\text{spin}}[V]$ on the spin bias $V$ for $V\ll J_{K}$.}\label{Sachdev_Current_01}
\end{figure}

To determine the behavior of $I_{\text{spin}}[V]$ for both Dirac and chiral KSL, we consider the contribution of all energy bands of the slab Hamiltonian for a system with $L_{\parallel}=20$ unit cells separating the two edges along the $\hat{\bm{y}}$ direction. In addition, we substitute the momentum integrals appearing in Eq. \eqref{Eq_SachdevDSSF02} by a sum over a mesh with $200\times 200$ points. In Fig. \ref{Sachdev_Current_01}, we show the dependence of the calculated spin current $I_{\text{spin}}[V]$ as a function of both the spin bias and the chiral interaction. We find that $I_{\text{spin}}[V]$ for the Dirac KSL depends on the spin bias as a  power-law $I_{\text{spin}}[V]\propto V^{\alpha}$ with the exponent $\alpha_{\text{Dirac}}\approx 1$. This linear dependence for arbitrarily small $V$ stems from the delta-function contribution to the density of the states associated with the zero-energy edge states. On the other hand, for the chiral KSL, we also find that the spin current $I_{\text{spin}}[V]$ for the chiral KSL depends approximately on the spin bias $V$ as a power-law for $V\ll J_{K}$. However, the exponent in this latter case is given by $\alpha_{\text{Chiral}}\approx 3$. As a consequence, this spin current becomes suppressed  within the regime of small spin bias more rapidly than the one for the Dirac KSL. It is also interesting to note that the exponent for the chiral KSL agrees with the exponent $\alpha=3$ found for a 1D spinon Fermi sea  in Ref. \cite{Chen-PRB(2013)} and can be associated with the gapless edge state. In the absence of a gapless edge mode, the spin current would vanish at zero temperature for  a spin bias  $V$ below the bulk spin gap. Another important point that we would like to emphasize here is that, in the limit of large spin bias (\emph{i.e.}, for $V\gtrsim J_{K}$), the spin currents that appear in both KSL states become asymptotically described by $I_{\text{spin}}[V]\propto V$.

\section{Longitudinal spin Seebeck effect driven by Majorana fermions}\label{Section_SIII}

The longitudinal spin Seebeck (LSS) effect refers to the generation of spin current flowing from a magnetic insulator to a metal, when they are subjected to the action of a static magnetic field and a temperature gradient across their interface \cite{Saitoh-NP(2017)}. According to linear response theory, based on the Keldysh formalism \cite{Maekawa-PRB(2011),*Maekawa-JKPS(2013)}, this spin current can be calculated by the formula 
\begin{align}\label{Eq_SeebeckI01}
I_{\text{spin}}[\bm{B}]=&\frac{4J^{2}A_{\perp}}{\sqrt{2}}\int^{\infty}_{-\infty}\frac{d\omega}{2\pi}\int\frac{d^{d-1}q_{\perp}}{(2\pi)^{d-1}}[\mathit{n}_{B}(T_{m})-\mathit{n}_{B}(T_{s})]\nonumber\\
&\times\chi_{-+}(\bm{q}_{\perp},\omega)S_{-+}(\bm{q}_{\perp},\omega,\bm{B}).
\end{align}
Here, $A_{\perp}$ and $J$ stand for the $(d-1)$-dimensional interface area and the exchange coupling of the metal with the magnetic insulator, $\mathit{n}_{B}(T)=1/(e^{\beta\omega}-1)$ is the Bose-Einstein distribution function, and $T_{m}$ and $T_{s}$ refer, respectively, to the temperature in the metal and the magnetic insulator. In this formula, the two functions $\chi_{-+}(\bm{q}_{\perp},\omega)$ and $S_{-+}(\bm{q}_{\perp},\omega,\bm{B})$ represent, respectively, the spin susceptibility of the metal and the DSSF of the magnetic insulator in the presence of a magnetic field $\bm{B}$. 

\begin{figure}[t]
\centering
\includegraphics[width=0.95\linewidth]{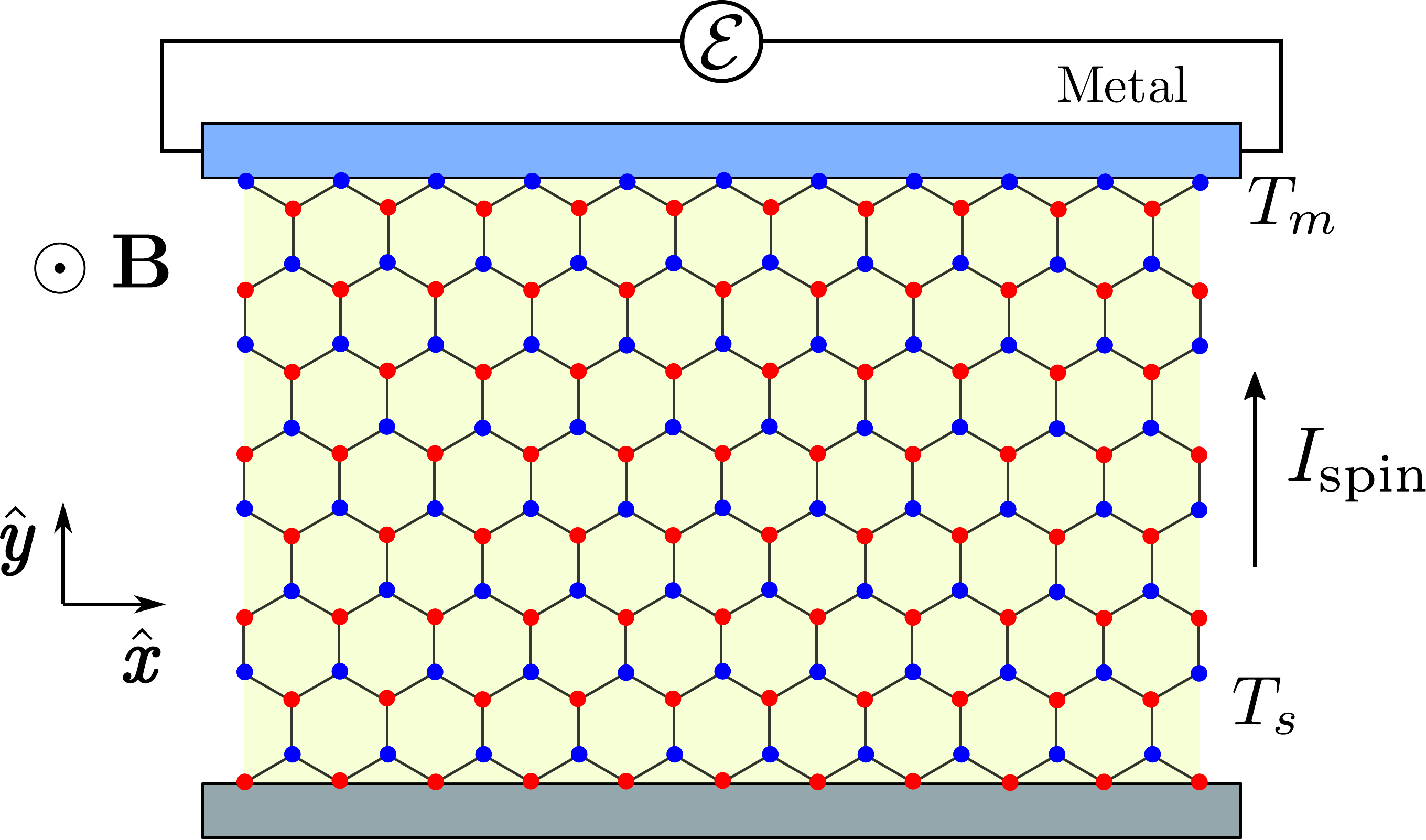}
\caption{(Color online) Illustration of the experimental setup for the observation of the LSS effect on an SU(2)-symmetric KSL system described by the Yao-Lee model. In order for this effect to occur, the spin system and a metal are placed in a region with finite magnetic field $\bm{B}=B_{z}\hat{\bm{z}}$ and kept at different temperatures $T_{s}$ and $T_{m}$, respectively. As a result, a finite spin current $I_{\text{spin}}$ can be driven at their interface and transformed in the metal into an electric current with an associated voltage $\mathcal{E}$ by means of the inverse spin Hall effect.}\label{SM_Seebeck_Effect_Setup}
\end{figure}

As the formula in Eq. \eqref{Eq_SeebeckI01} holds even for magnetic insulators lacking any kind of spontaneous-symmetry-breaking order, we shall apply it to the SU(2)-symmetric Yao-Lee model studied here. Henceforth, we will consider the metal as a clean 2D Fermi liquid system and approximate $\chi_{-+}(\bm{q}_{\perp},\omega)$ by a momentum-independent spin-diffusion type function \cite{Woelfle-PRB(1972)}. Therefore, we write it as $\chi_{-+}(\bm{q}_{\perp},\omega)=\chi_{0}\tau_{s}\omega/(1+\tau^{2}_{s}\omega^{2})$, where $\chi_{0}$ is the static spin susceptibility and $\tau_{s}$ corresponds to the spin relaxation time. On the other hand, we will evaluate $S_{-+}(\bm{q}_{\perp},\omega,\bm{B})$ [see Eq. \eqref{Eq_SachdevDSSF01}] by taking into account the  Hamiltonian in Eq. \eqref{Eq_YaoLee_Ham} for the slab geometry with zigzag surfaces, when the KSL system is placed in a region with a finite magnetic field $\bm{B}=B_{z}\hat{\bm{z}}$ perpendicular to the plane   as schematically depicted in Fig. \ref{SM_Seebeck_Effect_Setup}. 

The expression for the spin current $I_{\text{spin}}[B_{z}]$ will be further simplified here by expanding $\mathit{n}_{B}(T_{m})-\mathit{n}_{B}(T_{s})$ around $T=(T_{m}+T_{s})/2$ up to first order in $\Delta T= T_{m}-T_{s}$, \textit{i.e.}, the temperature gradient across  the interface. This leads to the following expression
\begin{align}\label{Eq_SeebeckI02}
I_{\text{spin}}[B_{z}]=&\frac{J^{2}L_{\perp}\chi_{0}\tau_{s}}{2\sqrt{2}\pi}\frac{\Delta T}{T^2}\int^{\infty}_{-\infty}\frac{d\omega}{\sinh^{2}[\omega/(2T)]}\frac{\omega^{2}}{1+\tau^{2}_{s}\omega^{2}}\nonumber\\
&\times\int\frac{dq_{\perp}}{2\pi}S_{-+}(\bm{q}_{\perp},\omega,B_{z}),
\end{align}
where we have substituted $A_{\perp}$ by the 1D interface area $L_{\perp}$ defined before.  According to this last result, the spin current $I_{\text{spin}}[B_{z}]$ is identically zero when the integrated DSSF $S_{-+}(\omega,B_{z})=\int\frac{dq_{\perp}}{2\pi}S_{-+}(\bm{q}_{\perp},\omega,B_{z})$ is an odd function of frequency. In the case of zero magnetic field, one may easily obtain by means of the Lehmann representation that the DSSF for the Yao-Lee model obeys $S_{-+}(\omega,B_{z}=0)=-S_{-+}(-\omega,B_{z}=0)$, even in the absence of time-reversal symmetry caused by a finite chiral interaction $J_{\chi}$. As a result, a non-zero spin current driven by the Majorana fermions in KSL and due to the LSS effect can only occur if the spin system is subjected to both a magnetic field and a temperature gradient at the interface with the metal. 

Once more, in order to determine $S_{-+}(\bm{q}_{\perp},\omega,B_{z})$ and subsequently $I_{\text{spin}}[B_{z}]$, we begin by calculating the Matsubara spin correlation function $X_{-+}(\bm{q}_{\perp},i\omega_{n},B_{z})$ defined by Eq. \eqref{Eq_Mats01}. As shown in Fig. \ref{SM_Seebeck_Effect_Setup}, we also impose here periodic boundary conditions on the spin system   along the $\hat{\bm{x}}$ direction, such that $\bm{q}_{\perp}=q_{\perp}\hat{\bm{x}}$. In order to find $X_{-+}(\bm{q}_{\perp},i\omega_{n},B_{z})$, we first evaluate the spin correlation function $\langle T_{\tau}[\sigma^{-}_{b}(\bm{R},\tau)\sigma^{+}_{b'}(\bm{R}',0)]\rangle$, as was done in the previous section. However, there is one important difference in the present case. Since the system is subjected to a finite magnetic field $\bm{B}=B_{z}\hat{\bm{z}}$, the expectation values of products of the Majorana fermions of $x$ and $y$ species are non-zero. Taking this fact into account, we find 
\begin{widetext}
\begin{align}\label{Eq_Mats03}
X_{-+}(q_{\perp},i\omega_{n},B_{z})=&\frac{1}{4L_{\perp}}\sum\limits^{2L_{\parallel}}_{\ell_{1}=1}\sum\limits^{2L_{\parallel}}_{\ell_{2}=1}\sum\limits^{}_{\nu=\pm}\sum\limits^{}_{k_{\perp}}\Bigl\vert\sum\limits^{}_{\ell\in\partial\mathcal{A}}U_{\ell,\ell_{1}}(k_{\perp})U_{\ell,\ell_{2}}(-k_{\perp}-q_{\perp})\Bigr\vert^{2}\bigl\{\mathit{n}_{F}[-E_{\ell_{1},\nu}(k_{\perp})]\mathit{n}_{F}[-E_{\ell_{2},z}(-k_{\perp}-q_{\perp})]\nonumber\\
&+\nu\mathit{n}_{F}[-E_{\ell_{1},\nu}(k_{\perp})]\mathit{n}_{F}[-E_{\ell_{2},z}(-k_{\perp}-q_{\perp})]\bigr\}\frac{\exp\{-\beta_{s}[E_{\ell_{1},\nu}(k_{\perp})+E_{\ell_{2},z}(-k_{\perp}-q_{\perp})]\}-1}{i\omega_{n}-E_{\ell_{1},\nu}(k_{\perp})-E_{\ell_{2},z}(-k_{\perp}-q_{\perp})},
\end{align}
\end{widetext}
where $E_{\ell,\nu}(k_{\perp})=E_{\ell,z}(k_{\perp})-\nu B_{z}/2$ are the energy dispersions of the slab Hamiltonian for a finite magnetic field. We note here that the second term on the r.h.s of Eq. \eqref{Eq_Mats03} has a finite contribution just in the situation where the   magnetic field $B_{z}$ is different from zero. On the other hand, the first term on the r.h.s. coincides with  the correlation function in Eq. \eqref{Eq_Mats02} when $B_{z}=0$.

From the Matsubara correlation function $X_{-+}(q_{\perp},i\omega_{n},B_{z})$, we compute the DSSF for a finite magnetic field   using Eq. (\ref{DSSF}). As a consequence, we obtain that the integrated DSSF $S_{-+}(\omega,B_{z})$ can be conveniently written as 
\begin{equation}
S_{-+}(\omega,B_{z})=S_{-+,I}(\omega,B_{z})+S_{-+,II}(\omega,B_{z}),
\end{equation}
where $S_{-+,I}(\omega,B_{z})$ and $S_{-+,II}(\omega,B_{z})$ are defined, respectively, as the momentum-integrated imaginary part of the first and second term of $X_{-+}(q_{\perp},i\omega_{n},B_{z})$ in Eq. (\ref{Eq_Mats03}) after taking $i\omega_{n}\rightarrow\omega+i\eta$. 

Before evaluating the spin current $I_{\text{spin}}[B_{z}]$, it is worth determining the parity of $S_{-+,I}(\omega,B_{z})$ and $S_{-+,II}(\omega,B_{z})$ under the change of $\omega\rightarrow-\omega$. By invoking the particle-hole symmetry properties of the eigenvalues and eigenvectors of the slab Hamiltonian (see Appendix for details), we obtain
\begin{align}
S_{-+,I}(-\omega,B_{z})=&-S_{-+,I}(\omega,B_{z}),\\
S_{-+,II}(-\omega,B_{z})=&\; S_{-+,II}(\omega,B_{z}).
\end{align}
As a result, only the frequency integration over $S_{-+,II}(\omega,B_{z})$ in Eq. \eqref{Eq_SeebeckI02} will give a finite contribution to the spin current $I_{\text{spin}}[B_{z}]$. In fact, it evaluates to
\begin{align}
I_{\text{spin}}[B_{z}]=&\frac{L_{\perp}J^{2}\chi_{0}\tau_{s}}{\sqrt{2}\pi}\frac{\Delta T}{T^2}\int^{\infty}_{0}\frac{d\omega}{\sinh^{2}[\omega/(2T)]}\frac{\omega^{2}}{1+\tau^{2}_{s}\omega^{2}}\nonumber\\
&\times S_{-+,II}(\omega,B_{z}).
\end{align}
In addition, we restrict ourselves to the limit $T_{s}\rightarrow 0$, in which we can neglect thermal fluctuations of the $\mathbb{Z}_{2}$ gauge field and determine  $I_{\text{spin}}[B_{z}]$ from the spectrum of the $c^\gamma$ Majorana fermions in the uniform background  $u_{\langle jk\rangle_{\alpha}}=1\; (-1)$ for a site of the sublattice $j=A\; (B)$. In this case, we have  
\begin{align}
I_{\text{spin}}[B_{z}]\stackrel{T_{s}\rightarrow 0}{=}&\frac{4J^{2}L_{\perp}\chi_{0}\tau_{s}}{\sqrt{2}\pi T_{m}}\int^{\infty}_{0}\frac{d\omega}{\sinh^{2}(\omega/T_{m})}\frac{\omega^{2}}{1+\tau^{2}_{s}\omega^{2}}\nonumber\\
&\times S_{-+,II}(\omega,B_{z}),
\end{align}
where the zero-temperature limit of the integrated DSSF $S_{-+,II}(\omega,B_{z})$ yields 
\begin{widetext}
\begin{align}
S_{-+,II}(\omega,B_{z})\stackrel{T_{s}\rightarrow 0}{=}&\; \frac{\pi}{4}\sum\limits^{2L_{\parallel}}_{\ell_{1}=1}\sum\limits^{2L_{\parallel}}_{\ell_{2}=1}\sum\limits^{}_{\nu=\pm}\int\frac{dq_{\perp}}{2\pi}\frac{dk_{\perp}}{2\pi}\nu\Bigl\vert\sum\limits^{}_{\ell\in\partial\mathcal{A}}U_{\ell,\ell_{1}}(k_{\perp})U_{\ell,\ell_{2}}(-k_{\perp}-q_{\perp})\Bigr\vert^{2}\Theta[E_{\ell_{1},\nu}(k_{\perp})]\nonumber\\
&\times\Theta[E_{\ell_{2},z}(-k_{\perp}-q_{\perp})]\delta[\omega-E_{\ell_{1},\nu}(k_{\perp})-E_{\ell_{2},z}(-k_{\perp}-q_{\perp})].
\end{align}
\end{widetext}
In contrast to the DSSF in Eq. \eqref{Eq_SachdevDSSF02}, we will now have to consider both the positive and negative energy states $E_{\ell,z}(k_{\perp})$ of the slab Hamiltonian in order to evaluate $S_{-+,II}(\omega,B_{z})$. The reason  is  that  the magnetic field here mixes the particle-hole states of the spin system. 

\begin{figure}[t]
\centering
\includegraphics[width=1.0\linewidth]{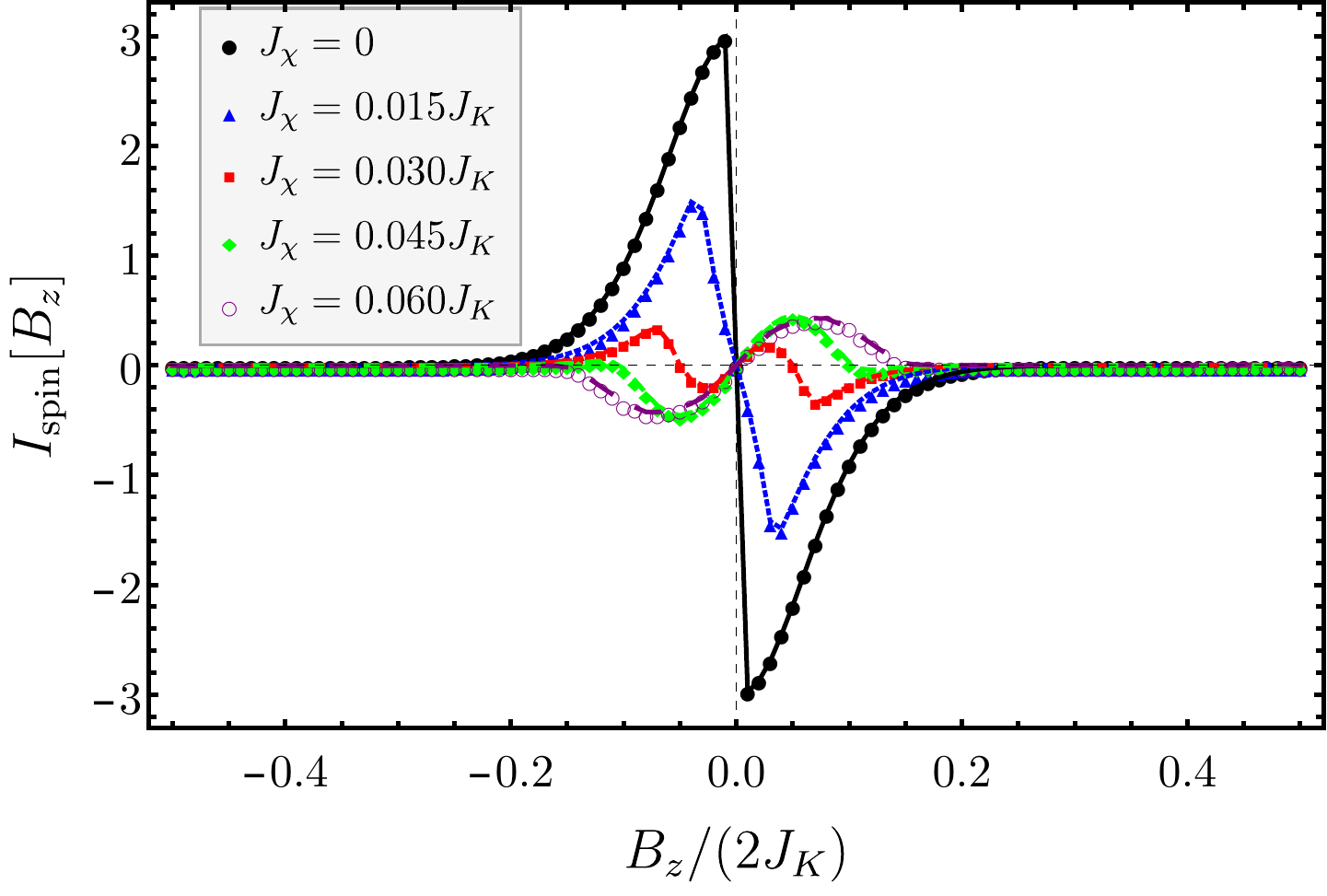}
\caption{(Color online) (Color online) Spin current $I_{\text{spin}}[B_{z}]$ generated by the LSS effect in a system consisting of a normal metal in contact with a Dirac or chiral KSL described by the Yao-Lee model. For magnetic fields $|B_{z}|\ll J_{K}$, $I_{\text{spin}}[B_{z}]$ exhibits a continuous change in the flow direction as the chiral interaction $J_{\chi}$ varies from zero to positive finite values. This feature can be used to identify the topological nature of KSLs, since the most relevant contribution to the spin current comes from the edge states. Here, $I_{\text{spin}}[B_{z}]$ is given in units of $2\sqrt{2}J^{2}L_{\perp}\chi_{0}J_{K}$, the metal spin relaxation is $\tau_{s}=1/(2J_{K})$, and metal and KSL temperatures are set to $T_{m}=0.1J_{K}$ and $T_{s}=0$, respectively.}\label{Seebeck_Current}
\end{figure}

We will focus here on determining the behavior of $I_{\text{spin}}[B_{z}]$ as a function of the magnetic field $B_{z}$ and the exchange interactions of the Yao-Lee model. In order to do that, we evaluate $S_{-+,II}(\omega,B_{z})$ by substituting the momentum integrals on its right-hand side by a sum over a mesh with $200\times 200$ points. Within this approximation, we show in Fig. \ref{Seebeck_Current} the behavior of $I_{\text{spin}}[B_{z}]$ as a function of both $B_{z}$ and $J_{\chi}$, and fixed metal temperature $T_{m}=0.1J_{K}$. We can observe that the spin current $I_{\text{spin}}[B_{z}]$  exhibits a continuous sign-change for fixed magnetic field $|B_{z}|\ll J_{K}$, as  the chiral interaction $J_{\chi}$  becomes finite, \textit{i.e.}, the spin current for the Dirac and chiral KSL states flows in opposite directions above a critical value $J^{c}_{\chi}$ of the chiral interaction. Additionally,  we note that the spin current for the former KSL state  is finite for infinitesimally small $B_{z}$, while for the latter case it is clearly suppressed as a linear function of $B_{z}$. The jump  in the spin current at $B_z=0$ for the Dirac KSL is a signature of the zero-energy edge states, which give a delta-function contribution to the density of states. Lastly, we point out that, in the limit of high magnetic  fields,  the spin currents vanish asymptotically  for both KSL states. 

The physical properties of the spin current $I_{\text{spin}}[B_{z}]$ investigated here are   therefore different from the mechanism of the LSS effect that appears in the one-dimensional QSL Sr$_{2}$CuO$_{3}$ in Ref. \cite{Saitoh-NP(2017)}, in which the low-energy spinons emerging in an antiferromagnetic Heisenberg chain are described by the TL theory. In that case, the spin current due to the LSS effect is non-zero only when the particle-hole symmetry is broken by quadratic terms in the spinon dispersion. In the same work  (Ref. \cite{Saitoh-NP(2017)}), the sign difference in the Seebeck response of different insulating magnetic materials was argued to be related to either the existence or non-existence  of long-range magnetic order in the system. Interestingly, according to our present results, the sign of the $I_{\text{spin}}[B_{z}]$ for a fixed magnetic field can also be tuned by the edge states  by means of a chiral interaction in systems  composed only of particle-hole symmetric Majorana fermion excitations.

\section{Conclusions}\label{Section_SIV}

In summary, we have analyzed both the edge magnetism and the spin transport properties of an SU(2)-symmetric spin-$1/2$ system described by the Yao-Lee model, in which the spin degrees of freedom fractionalize and give rise to either a Dirac or chiral KSL with Majorana fermionic excitations. We have shown that the edge magnetization for the Dirac KSL possesses a nonlocal response when the spin system is probed by local magnetic fields. These features are a consequence of both the long-range correlations between the spin-$1/2$ moments and the zero-energy edge bands of the 2D spin system with a zigzag edge. 

We have also shown that the application of a spin bias at the interface of the spin system with a normal metal generates a spin current into the KSL states depending as a  power-law on the spin bias in the zero-temperature limit. As the spin current response depends on the dynamical spin structure factor at the boundary of the spin system, it is sensitive to both bulk and  edge states in the spectrum  of the KSL. As a result,  the spin current measured in this setup can be used to identify the  nodal structures formed by the Majorana excitations  in KSLs with SU(2) symmetries, similarly to what has been discussed for other types of QSLs in the literature \cite{Sachdev-PRB(2015),Chen-PRB(2013)}. Moreover, we have also predicted here that a spin current can be controlled at the interface of the KSL with a metal by the LSS effect, despite the particle-hole symmetry of the Majorana excitations in this case.

Lastly, we should also comment on the stability of the spin currents calculated here. In real materials, the SU(2) spin symmetry is broken by, \textit{e.g.}, Rashba or hyperfine interactions. However, as long as the SU(2)-symmetry breaking interactions are weak, the spin current can propagate over long distances into the QSL.  We hope that our results   will contribute to stimulate further activities in the experimental pursuit of novel KSL materials, particularly those with fluctuating orbital degrees of freedom that might realize  SU(2)-symmetric KSLs as described by the Yao-Lee model.  

\section*{Acknowledgments}\label{Section_SV}

Two of us (V.S.deC. and E.M.) would like to thank C. Farias for stimulating discussions. They also thank the financial support from FAPESP under Grant No. 2016/05069-7, and the kindly hospitality of the International Institute of Physics (IIP), in Natal, where part of this work was done; IIP is  supported by the Brazilian ministries MEC and MCTIC. H.F. acknowledges funding from CNPq under Grant No. 405584/2016-4. 


\setcounter{equation}{0}
\setcounter{figure}{0}
\renewcommand{\theequation}{A\arabic{equation}}
\renewcommand{\thefigure}{A\arabic{figure}}

\section*{Appendix: Slab Hamiltonian for a zig-zag surface}\label{Section_SVI}

In order to obtain the Yao-Lee Hamiltonian $\mathcal{H}$ on the honeycomb lattice for a slab geometry with periodic boundary conditions along the $\hat{\bm{x}}$ direction and zigzag surfaces along $\hat{\bm{y}}$ direction, we initially fix the $\mathbb{Z}_{2}$ bond variables in  Eq. \eqref{Eq_MajHam} according to $u_{\langle jk\rangle_{\alpha}}=1\;(-1)$ for a site $j$ in the $A\;(B)$ sublattice, and define the partial Fourier transform 
\begin{equation}\label{Eq_PartFourier}
c^{\alpha}_{b}(x,y)=\frac{1}{\sqrt{L_{\perp}}}\sum\limits^{}_{k_{\perp}}e^{-ik_{\perp}\cdot x}c^{\alpha}_{b}(k_{\perp},y)
\end{equation}
for the Majorana fermion operators. Here, $b\in\{A,B\}$ is the sublattice index, $L_{\perp}$ is the number of unit cells on the edge along the $\hat{\bm{x}}$ direction, $k_{\perp}$ is the wave vector related to it, and $y=1,\ldots,L_{\parallel}$ represents the number of unit cells along the $\hat{\bm{y}}$ direction. Note that the $(k_{\perp},y)$-dependent Majorana fermion operators $c^{\alpha}_{b}(k_{\perp},y)$ in Eq. \eqref{Eq_PartFourier} are related to their Hermitian conjugates by the relation $c^{\alpha}_{b}(-k_{\perp},y)=[c^{\alpha}_{b}(k_{\perp},y)]^{\dagger}$. By considering the coupling of the spin-$1/2$ moments to the magnetic field $\bm{B}=B_z\hat{\bm{z}}$, the Yao-Lee Hamiltonian becomes  
\begin{align}
\mathcal{H}=&\; J_{K}\sum\limits^{}_{\langle jk\rangle_{\alpha}}\sum\limits^{}_{\gamma=x,y,z}\hat{u}_{\langle jk\rangle_{\alpha}}ic^{\gamma}_{j}c^{\gamma}_{k}\nonumber\\
&+J_{\chi}\sum\limits^{}_{\langle jk\rangle_{\alpha}}\sum\limits^{}_{\langle kl\rangle_{\beta}}\sum\limits^{}_{\gamma=x,y,z}\hat{u}_{\langle jk\rangle_{\alpha}}\hat{u}_{\langle kl\rangle_{\beta}}ic^{\gamma}_{j}c^{\gamma}_{l}\nonumber\\
&+B_{z}\sum\limits^{}_{j}ic^{x}_{j}c^{y}_{j}.
\end{align}
In that case, by making use of the partial Fourier transform defined above, we obtain that $\mathcal{H}$ evaluates to
\begin{equation}\label{Eq_SlabHam}
\mathcal{H}=\hspace{-0.2cm}\sum\limits^{}_{k_{\perp},\alpha}C^{\alpha\dagger}(k_{\perp})\mathcal{H}(k_{\perp})C^{\alpha}(k_{\perp})+iB_{z}\sum\limits^{}_{k_{\perp}}C^{x\dagger}(k_{\perp})C^{y}(k_{\perp}),
\end{equation}
where $C^{\alpha}(k_{\perp})\equiv[c^{\alpha}_{A}(k_{\perp},1),c^{\alpha}_{B}(k_{\perp},1),\ldots,c^{\alpha}_{B}(k_{\perp},L_{\parallel})]^{T}$ is a $2L_{\parallel}$-dimensional spinor and $\mathcal{H}(k_{\perp})$ is given by
\begin{align}
&\mathcal{H}(k_{\perp})\nonumber\\
&=
\begin{pmatrix}
\mathcal{A}(k_{\perp}) & \mathcal{B}^{\dagger}(k_{\perp}) & 0                       & 0                       & \dots  & 0 \\
\mathcal{B}(k_{\perp}) & \mathcal{A}(k_{\perp})           & \mathcal{B}^{\dagger}(k_{\perp}) & 0                       & \dots             & 0 \\
0             & \mathcal{B}(k_{\perp})           & \mathcal{A}(k_{\perp})           & \mathcal{B}^{\dagger}(k_{\perp}) &  \ddots  & \vdots \\
0             & 0                       & \mathcal{B}(k_{\perp})           & \ddots                  & \ddots &  0  \\
\vdots        & \vdots                  & \ddots                  & \ddots                  & \mathcal{A}(k_{\perp})  & \mathcal{B}^{\dagger}(k_{\perp}) \\
0             & 0                       &  \dots                  & 0                       & \mathcal{B}(k_{\perp})  & \mathcal{A}(k_{\perp})
\end{pmatrix},
\end{align}
with $\mathcal{A}(k_{\perp})$ and $\mathcal{B}(k_{\perp})$ being square matrices defined according to
\begin{equation}
\mathcal{A}(k_{\perp})\equiv
\begin{pmatrix}
J_{\chi}\sin(k_{\perp}a)  &   iJ_{K}/2   \\
-iJ_{K}/2 & -J_{\chi}\sin(k_{\perp}a)
\end{pmatrix},
\end{equation}
and 
\begin{equation}
\mathcal{B}(k_{\perp})\equiv
\begin{pmatrix}
-J_{\chi}\sin(k_{\perp}a) & iJ_{K}\cos(k_{\perp}a/2) \\
0 & J_{\chi}\sin(k_{\perp}a) 
\end{pmatrix}.
\end{equation}

By looking at the structure of $\mathcal{H}(k_{\perp})$, it is straightforward to verify that this matrix is skew-Hermitian, \textit{i.e.}, under the transpose operation, it transforms as $\mathcal{H}(k_{\perp})=-\mathcal{H}^{T}(-k_{\perp})$. Consequently, by defining a band index $\ell$, one may show that the entries of the eigenvectors $|U_{\ell}(k_{\perp})\rangle$ and the eigenvalues $E_{\ell}(k_{\perp})$ of $\mathcal{H}(k_{\perp})$ are characterized by the relations $U_{\ell,L_{\parallel}+\ell'}(k_{\perp})=U^{*}_{\ell,\ell'}(-k_{\perp})$ and $E_{L_{\parallel}+\ell}(k_{\perp})=-E_{\ell}(-k_{\perp})$, where the eigenvalues are distributed according to $E_{1}(k_{\perp})\leq E_{2}(k_{\perp})\leq\cdots\leq E_{L_{\parallel}}(k_{\perp})$ (see Fig. \ref{Edge_States_SI}). This allows one to rewrite each component of the Majorana spinor $C^{\alpha}(k_{\perp})$ as
\begin{equation}
C^{\alpha}_{\ell}(k_{\perp})=\sum\limits^{L_{\parallel}}_{\ell'=1}[U_{\ell,\ell'}(k_{\perp})\Psi^{\alpha}_{\ell'}(k_{\perp})+U^{*}_{\ell,\ell'}(-k_{\perp})\Psi^{\alpha\dagger}_{\ell'}(-k_{\perp})],
\end{equation}
where $\Psi^{\alpha\dagger}_{\ell}(k_{\perp})$ and $\Psi^{\alpha}_{\ell}(k_{\perp})$ are complex fermion operators obeying the anti-commutation relations $\{\Psi^{\alpha}_{\ell}(k_{\perp}),\Psi^{\alpha'\dagger}_{\ell'}(k_{\perp}')\}=\delta_{\alpha,\alpha'}\delta_{\ell,\ell'}\delta_{k_{\perp},k_{\perp}'}$. By inserting this last result into Eq. \eqref{Eq_SlabHam} and then performing the canonical transformation
\begin{equation}
\Psi^{\dagger}_{\ell,\pm}(k_{\perp})\equiv\frac{1}{\sqrt{2}}[\Psi^{x\dagger}_{\ell}(k_{\perp})\pm i\Psi^{y\dagger}_{\ell}(k_{\perp})],
\end{equation}
we obtain the following diagonal Hamiltonian
\begin{align}
\mathcal{H}=&\sum\limits^{L_{\parallel}}_{\ell=1}\sum\limits^{}_{k_{\perp}}\sum\limits^{}_{\nu=\pm}E_{\ell,\nu}(k_{\perp})\biggl[\Psi^{\dagger}_{\ell,\nu}(k_{\perp})\Psi_{\ell,\nu}(k_{\perp})-\frac{1}{2}\biggr]\nonumber\\
&+\sum\limits^{L_{\parallel}}_{\ell=1}\sum\limits^{}_{k_{\perp}}E_{\ell,z}(k_{\perp})\biggl[\Psi^{z\dagger}_{\ell}(k_{\perp})\Psi^{z}_{\ell}(k_{\perp})-\frac{1}{2}\biggr],
\end{align}
where $E_{\ell,\nu}(k_{\perp})\equiv E_{\ell,z}(k_{\perp})-\nu B_{z}$ and $E_{\ell,z}(k_{\perp})\equiv 2E_{\ell}(k_{\perp})$ are the energy dispersions of the Majorana fermions in the KSL for a finite magnetic field.


%

\end{document}